\documentclass[aps,twocolumn,amsmath,amssymb,reprint,floatfix,superscriptaddress]{revtex4-2}
\usepackage{amsmath,amssymb,bm,graphicx}
\usepackage{times}

\usepackage{hyperref}
\hypersetup{
colorlinks=true,
linkcolor=blue,           % color of internal links
citecolor=blue,           % color of links to bibliography
filecolor=magenta,        % color of file links
urlcolor=cyan
}

\newcommand{\papertitle}{The non-reciprocal Dicke model}

\begin{document}

\title{\papertitle}
\author{Ezequiel~I.~Rodr\'iguez Chiacchio}
\affiliation{Entropica Labs, 186B Telok Ayer Street 068632, Singapore}
\author{Andreas~Nunnenkamp}
\email{andreas.nunnenkamp@univie.ac.at}
\affiliation{Faculty of Physics, University of Vienna, Boltzmanngasse 5, 1090 Vienna, Austria}
\author{Matteo~Brunelli}
\email{matteo.brunelli@unibas.ch}
\affiliation{Department of Physics, University of Basel, Klingelbergstrasse 82, 4056 Basel, Switzerland}

%\date{\today}

\begin{abstract}
We investigate the physics of an open two-component Dicke model, where the light field mediates non-reciprocal interactions between two spin species. We show that the model, which we dub non-reciprocal Dicke model, exhibits a discrete parity-time ($\mathcal{PT}$) symmetry and we characterize the emergence of a non-stationary phase, so far explained in terms of dissipation-induced instability, as spontaneous breaking of $\mathcal{PT}$ symmetry. We further show that such $\mathcal{PT}$ symmetry breaking embodies an instance of a non-reciprocal phase transition, a concept recently introduced by Fruchart \emph{et al.} [Nature \textbf{592}, 363 (2021)]. Remarkably, the phase transition in our model does not necessitate the presence of any underlying broken symmetry or exceptional points in the spectrum, both believed to be essential requirements for non-reciprocal phase transitions. Our results establish driven-dissipative light-matter systems as a new avenue for exploring non-reciprocal phase transitions and contribute to the theory of non-reciprocal collective phenomena.
\end{abstract}

\maketitle

\textit{Introduction.---}
Newton's third law states that to every action there is always an equal and opposed  reaction~\cite{Newton1687}. 
% When interactions are mediated by non-equilibrium agents, this principle can be broken, giving rise to non-reciprocity. 
For non-equilibrium agents, this principle can be broken, giving rise to non-reciprocal interactions, namely, interactions which are not symmetric upon the exchange of agents.
This universal phenomenon has been observed in active matter~\cite{Uchida2010,Nagy2010,Yllanes2017,Lavergne2019,Saha2019,Shankar2022,Baconnier2022}, interface growth~\cite{Coullet1989,Pan1994}, neural systems~\cite{Montbrio2018}, and social dynamics~\cite{Hong2011}. Recently, Fruchart \emph{et al.}~\cite{Fruchart2021} made a seminal contribution to a general theory of non-reciprocal phase transitions (NRPTs), showing that non-reciprocal interactions can lead to non-stationary phases of matter and exceptional points (EPs). Their findings were illustrated in paradigmatic non-equilibrium models, such as flocking, pattern formation, and synchronization. In contrast, for engineered photonic and coupled light-matter systems, non-reciprocity has been mostly investigated in the context of signal transmission~\cite{Jalas2013,Fleury2014,Estep2014,Metelmann2015,Lodahl2017,Bernier2017,Verhagen2017,Miri2019}, with the role of many-body interactions left almost entirely unexplored~\cite{HanaiPRL2019}.

In this Letter, we provide the characterization of a NRPT in a many-body light-matter system by studying an open Dicke model~\cite{KeelingDicke} featuring two different spin species, with photons mediating non-reciprocal interactions between them. We show that non-reciprocal interactions account for the key features of the model, such as its steady-state phase diagram and dynamics, hence we dub it \textit{non-reciprocal Dicke model} (NRDM).
\begin{figure}[t]
\centering
\includegraphics[width=.9\columnwidth]{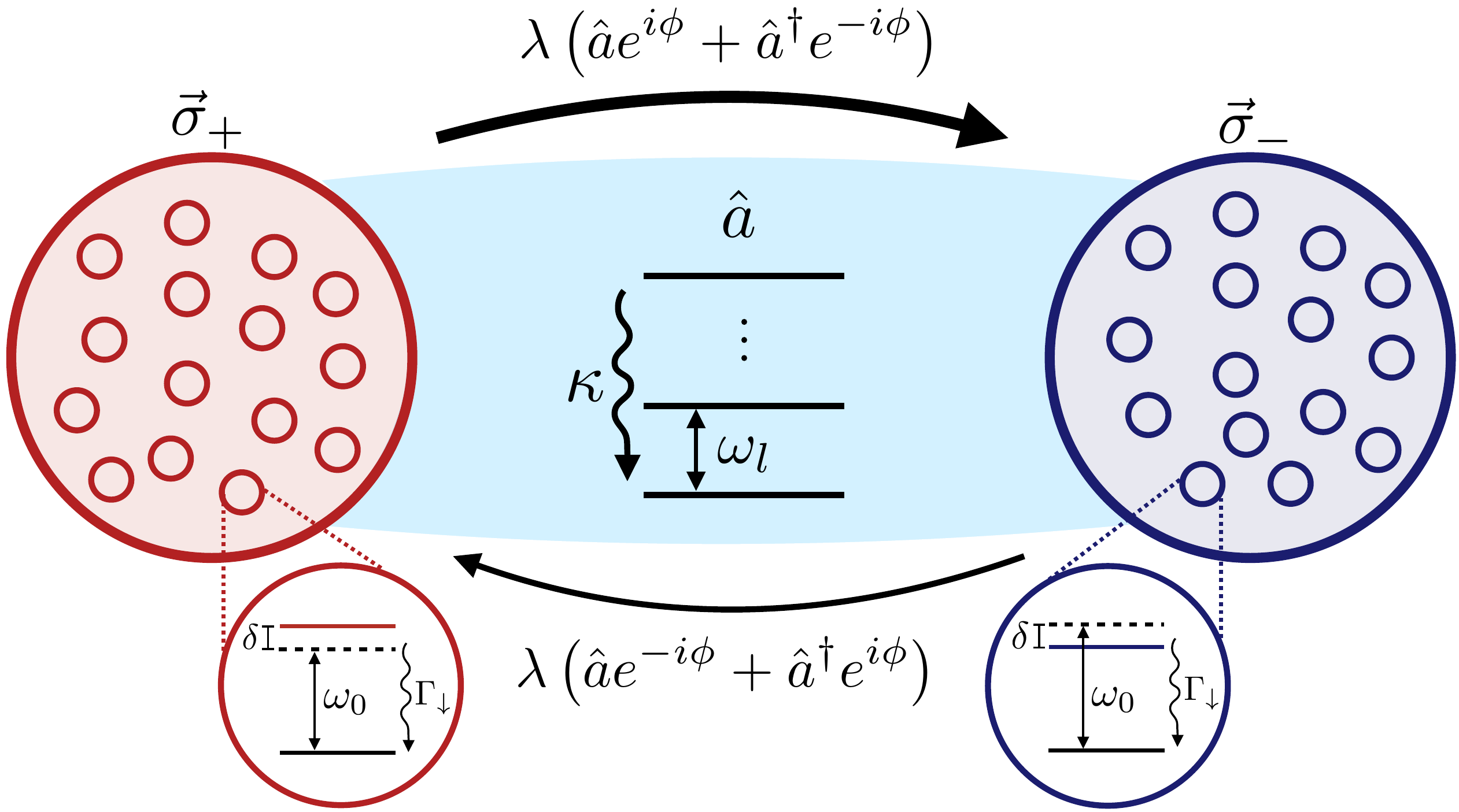}
\caption{\label{Model} \textbf{The non-reciprocal Dicke model}.
Two spin species, labelled $\pm$, each consisting of $N$ identical spins, interact with a light field $\hat{a}$ with frequency $\omega_{l}$ and decay rate $\kappa$, with coupling of modulus $\lambda$ and phase $\pm \phi$. This results in photon-mediated non-reciprocal interactions between the spins. The two species may have different frequencies $\omega_0 \pm \delta$ and be affected by spin relaxation at a rate $\Gamma_{\downarrow}$.}
\end{figure}

The most distinctive feature of the NRDM is arguably the emergence of a non-stationary phase, which we associate with the presence of a NRPT.
Going beyond the paradigm of Ref.~\cite{Fruchart2021}, we show that this transition takes place
(i) in the absence of any initial spontaneously broken symmetry and
(ii) in the absence of EPs, when non-reciprocal interactions are mediated by a \emph{dynamical} degree of freedom. Excitingly, we find that the NRPT is robust against spin frequency imbalance and spin decay. Our study suggests that NRPTs are a more general phenomenon than currently appreciated.

Specifically, we show that the NRDM is characterized by a parity-time ($\mathcal{PT}$) symmetry, in addition to the parity symmetry associated to the normal-to-superradiant phase transition, which is spontaneously broken in the NRPT. Remarkably, this $\mathcal{PT}$ symmetry breaking occurs at the level of the steady state and thus supersedes standard treatments restricted to transient growth or decay, as obtained from the eigenvalues of associated non-Hermitian Hamiltonians~\cite{Bender2007}.

The non-stationary phase of the NRDM has previously been characterized in terms of a dissipation-induced instability and chiral forces~\cite{Chiacchio2019b,Buca2019,Dogra2019} and even observed in a spinor Bose-Einstein condensate (BEC) in an optical cavity~\cite{Dogra2019}. Here we go beyond these considerations by identifying it as a symmetry broken phase belonging to the novel class of non-equilibrium NRPTs. We further uncover a rich non-stationary behavior, including frequency locking between the light and collective spin oscillations and the coexistence of superradiant and non-stationary behavior.

Our results bridge the fields of non-reciprocal critical phenomena and driven-dissipative light-matter systems and can be tested in state-of-the-art atom-cavity experiments \cite{Weiner2017, EsslingerSpin, Dogra2019, Kroeze2018, KeelingLevPRX, Periwal2021, Mivehvar2021}.

\begin{figure}[t]
\centering
\includegraphics[width=\columnwidth]{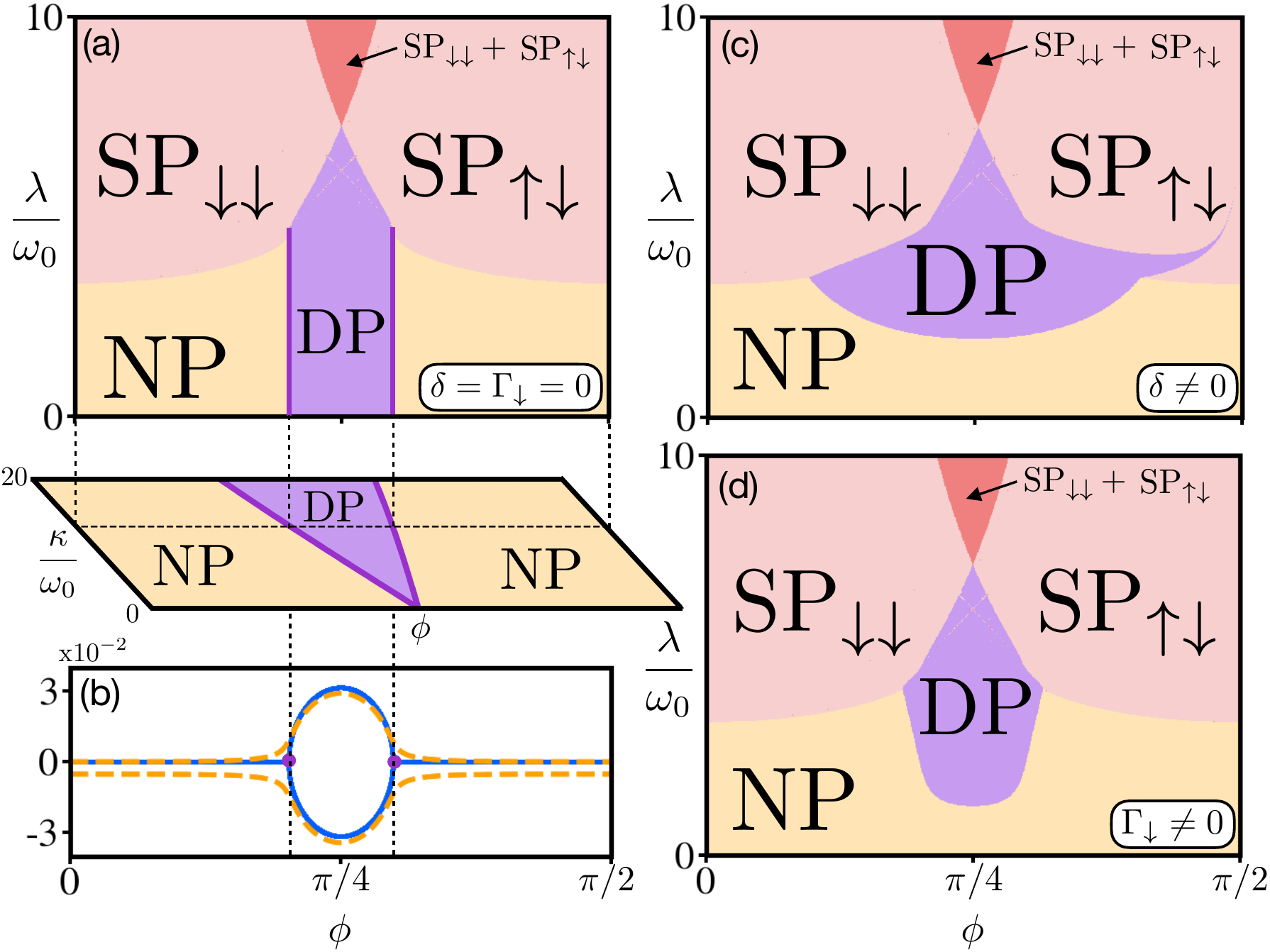}
\caption{\label{PD-Spec} \textbf{Effective non-reciprocal interactions.}
(a) Steady-state phase diagram as a function of coupling $\lambda$ and phase $\phi$, and of photon loss rate $\kappa$ and $\phi$ (out-of-plane view), in the limit of adiabatic elimination and for $\delta=\Gamma_{\downarrow} = 0$. Lines made of EPs are marked in violet. The solid blue line in (b) is the real part of the spectrum  around the NP for $\lambda = 2.5\omega_{0}$; violet dots mark EPs. \textbf{Photon-mediated non-reciprocal interactions.} Phase diagram in the case where non-reciprocal interactions are dynamically mediated by the light field for (c) $\delta = 0.05\omega_{0}$, $\Gamma_{\downarrow} = 0$, and (d) $\delta = 0$, $\Gamma_{\downarrow} = 0.02\omega_{0}$. The dashed yellow curve (b) is the real part of the spectrum for $\delta=\Gamma_{\downarrow} = 0$ and $\lambda = 2.5\omega_{0}$. Parameters: (a)-(d) $\omega_{l}  = 20\omega_{0}$, $\kappa = 12.5\omega_{0}$.
% \textbf{Effective non-reciprocal interactions.}
% (a) Steady-state phase diagram as a function of coupling $\lambda$ and phase $\phi$, and of photon loss rate $\kappa$ and $\phi$ (out-of-plane view), in the limit of adiabatic elimination. Lines made of exceptional points (EPs) are marked in violet. (b) Dynamical spectrum of the system around the NP solution, in the adiabatic regime (blue solid), and including photon fluctuations (yellow dashed), for $\lambda = 2.5\omega_{0}$; violet dots mark EPs. \textbf{Photon-mediated non-reciprocal interactions.} Phase diagram for (c) $\delta = 0.05\omega_{0}$, $\Gamma_{\downarrow} = 0$, and (d) $\delta = 0$, $\Gamma_{\downarrow} = 0.02\omega_{0}$, where non-reciprocal interactions are dynamically mediated by the light field. Parameters: (a)-(d) $\omega_{l}  = 20\omega_{0}$, $\kappa = 12.5\omega_{0}$.
}
\end{figure}

\textit{Model and symmetries.---}
The NRDM is an open Dicke model consisting of two different spin species and a light field featuring complex coupling amplitudes \cite{Dogra2019,Chiacchio2019b,Buca2019}, see Fig.~\ref{Model}.
The coherent dynamics are given by the Hamiltonian ($\hbar=1$)
\begin{equation}
\label{eq:er1}
\hat{H} = \hat{H}_{0} +  \frac{\lambda}{2\sqrt{N}}  \sum_{j=1}^{N} \sum_{m = \pm} \left(e^{-im\phi}\hat{a} + e^{im\phi}\hat{a}^{\dagger} \right) \hat{\sigma}^{x}_{j,m}\,,
\end{equation}
with $\hat{H}_{0} = \omega_{l} \hat{a}^{\dagger} \hat{a} + \frac{1}{2}\sum_{j,m}( \omega_0 + m\delta) \hat{\sigma}^{z}_{j,m}$, $\hat{a}$ the annihilation operator of the light field, $\omega_{l}$ the photon frequency, $\hat{\sigma}^{z}_{j,m}$ the $z$ Pauli matrix of the $j$-th spin of species $m=\pm$, $\omega_{0}$ the mean spin frequency, and $2\delta$ the frequency splitting between species.
The interaction is collective, i.e.,~of Dicke type, with $N$ the number of spins of each species.
Crucially, the light-matter coupling amplitudes are complex with modulus $\lambda$ and a species-dependent phase $\pm \phi$, which cannot be removed from the Hamiltonian by a gauge transformation.
Dissipative terms are incorporated in the model via a Lindblad master equation of the form $\dot{\hat{\rho}}=-i[\hat{H},\hat{\rho}] + \kappa \mathcal{D}[\hat{a}]\hat{\rho} + \Gamma_{\downarrow} \sum_{j,m} \mathcal{D}[\hat{\sigma}^{-}_{j,m}]\hat{\rho}$, where $\kappa$ is the photon loss rate, $\Gamma_{\downarrow}$ is the spin decay rate, with $\hat{\sigma}^{-} = (\hat{\sigma}^{x} - i\hat{\sigma}^{y})/2$ the spin lowering operator, and the dissipator defined as $\mathcal{D}[\hat{R}] \hat{\rho} = \hat{R}\hat{\rho}\hat{R}^{\dagger}-\frac{1}{2}\{\hat{R}^{\dagger}\hat{R},\hat{\rho}\}$.

In the thermodynamic limit, $N \rightarrow \infty$, the semi-classical equations of motion become exact
\begin{align}
\label{eq:er2}
&\dot{s}_{x,\pm} = -( \omega_0 \pm \delta) s_{y,\pm} - \frac{\Gamma_{\downarrow}}{2} s_{x,\pm} \nonumber \\
&\dot{s}_{y,\pm} = ( \omega_0 \pm \delta) s_{x,\pm} - \frac{\Gamma_{\downarrow}}{2} s_{y,\pm} - \frac{\lambda s_{z,\pm} }{\sqrt{N}} \left( \alpha e^{\mp i\phi} + \alpha^{*} e^{\pm i\phi}  \right) \nonumber \\
&\dot{s}_{z,\pm} = -\Gamma_{\downarrow} (s_{z,\pm} + 1) +  \frac{\lambda s_{y,\pm}}{\sqrt{N}} \left( \alpha e^{\mp i\phi} + \alpha^{*} e^{\pm i\phi}  \right) \nonumber \\
&\dot{\alpha} = -\left(i\omega_{l} + \frac{\kappa}{2} \right) \alpha - i\frac{\lambda\sqrt{N}}{2} \left( s_{x,+} e^{i\phi} + s_{x,-} e^{-i\phi} \right),
\end{align}
with $\alpha  = \langle \hat{a} \rangle$ and $s_{(x,y,z),\pm} = \langle \hat{\sigma}^{(x,y,z)}_{j,\pm} \rangle$.

As many Dicke models, the NRDM features a $\mathbb{Z}_{2}$ parity symmetry, $\nobreak{(\alpha, s_{x,\pm},s_{y,\pm}) \rightarrow -(\alpha, s_{x,\pm}, s_{y,\pm})}$. Spontaneous breaking of this symmetry leads to a superradiant phase transition \cite{HeppLieb,WangHioe,KeelingDicke}. This corresponds to a transition between a normal phase (NP), where the light field is empty and the spins point down, and a superradiant phase (SP), where the spins acquire a finite $x$-component and the light field is macroscopically populated.

The distinctive trait of the NRDM is that the light field mediates non-reciprocal interactions between the spin species. Their origin can be traced back to the joint presence of the energy non-conserving terms characteristic of the Dicke model, responsible for the gauge-invariant phase $2\phi$ in Eq.~\eqref{eq:er1}, and photon losses~\cite{SuppMatt}. Non-reciprocal interactions give rise to a region in the phase diagram displaying non-stationary steady-states, called the dynamical phase (DP)~\cite{Dogra2019, Chiacchio2019b, Buca2019}. Non-reciprocal interactions also affect the superradiant region, resulting in the emergence three different phases~\cite{Chiacchio2019b}: one in which the spins are almost aligned (SP$_{\downarrow\downarrow}$), another in which they are almost anti-aligned (SP$_{\uparrow\downarrow}$), and a third one corresponding to a coexistence region between the two.

From Eqs.~\eqref{eq:er2} we notice that the NRDM exhibits a second discrete symmetry, associated with the transformation $\nobreak{(s_{j,\pm},\phi) \rightarrow (s_{j,\mp},-\phi)}$  with $j=x,y,z$, which combines a parity transformation that swaps the two species with a change of sign of $\phi$. Since the phase $\phi$ acts as a synthetic magnetic flux, a change in sign describes the action of time reversal, so that the combined transformation is equivalent to $\mathcal{PT}$ symmetry. We stress that this is an exact symmetry of the nonlinear set of Eqs.~\eqref{eq:er2}, valid for all values of parameters. This has to be contrasted with standard treatments of $\mathcal{PT}$ symmetry in linear models, formulated in terms of non-Hermitian operators. A similar notion of $\mathcal{PT}$ symmetry has been studied in a  dimer with saturable loss and gain~\cite{Kepesidis2016}. $\mathcal{PT}$ symmetry has also been discussed at the level of Lindblad dynamics~\cite{Prosen2012,Huber2020,Nakanishi2022,Nakanishi2023}. We will show that, in the absence of explicit symmetry breaking terms, i.e., when $\delta=0$, the spontaneous breaking of $\mathcal{PT}$ symmetry is the hallmark of the NRPT, as it is spontaneously broken when entering the DP, but unbroken in the NP and SPs.

\textit{Non-reciprocal phase transition.---}
Our analysis begins with the steady-state phase diagram, which we obtain by setting Eqs.~\eqref{eq:er2} to zero, solving the set of algebraic equations, and performing linear stability analysis~\cite{SuppMatt}.
We first illustrate the NRPT in the regime $(\omega_{l},\kappa) \gg (\omega_{0},\delta,\Gamma_{\downarrow},\lambda)$, where the photonic degree of freedom can be adiabatically eliminated. We refer to this regime as effective non-reciprocal interactions, because non-reciprocity is encoded as effective asymmetrical coupling constants in the equations of motion for the two spin species~\cite{SuppMatt}. In this scenario we recover the treatment of Ref.~\cite{Fruchart2021}.
The steady-state phase diagram is shown in Fig.~\ref{PD-Spec}(a) for $\delta=\Gamma_{\downarrow}=0$. We observe the emergence of the DP in the central region. We characterize this phase exploiting the fact that the spectrum of the fluctuations around the NP can be computed analytically~\cite{Chiacchio2019b}. In Fig.~\ref{PD-Spec}(b), we show in blue the real part of the spectrum, associated with the growth rates of fluctuations, as a function of $\phi$. As the system crosses the NP-DP boundary, we observe the instability of the NP, heralded by the presence of EPs, resulting from non-reciprocity.
While not shown, we also observe eigenvector coalescing, a characteristic feature accompanying EPs. 
The occurrence of EP and a finite real part of the spectrum is a manifestation of $\mathcal{PT}$ symmetry breaking at the level of fluctuations~\cite{Fruchart2021}.
Note that for $\lambda = 0$, the system is trivially non-interacting, and thus remains in the NP.

In Fig.~\ref{PD-Spec}(a) we also show an out-of-plane section of the phase diagram  as a function of the photon loss rate $\kappa$ and phase $\phi$. Starting at $\kappa = 0$, where interactions are reciprocal, we observe how the DP emerges as soon as non-reciprocity ($\kappa \neq 0$) is turned on, with the phase boundaries made of EPs. Again, this agrees with the analysis in~\cite{Fruchart2021}: non-reciprocal interactions can open up a non-stationary region in the phase diagram, with lines of EPs present at the boundary. We thus conclude the NP-DP transition indeed corresponds to a NRPT. Nevertheless, there is a fundamental difference between the NRDM we consider and the models in~\cite{Fruchart2021}: the NRPT does not necessitate an underlying broken continuous symmetry. In fact, the NRPT here takes place in the absence of \textit{any} initially broken symmetry. 

We now take our investigation beyond adiabatic elimination and show that
NRPTs can take place in the absence of EPs in the spectrum, if the non-reciprocal interactions are mediated by a dynamical degree of freedom; we refer to this scenario as photon-mediated non-reciprocal interactions. We highlight two major findings. First, the DP remains present in the phase diagram, displaying robustness against finite frequency imbalance and spin decay, see Figs.~\ref{PD-Spec}(c), (d). In fact, for both $\delta\neq0$ (c) and $\Gamma_{\downarrow}\neq0$ (d), we find the NP to remain stable for a finite region, see also~\cite{SuppMatt} for the corresponding dynamical spectra.
We also note that for $\delta=\Gamma_{\downarrow}=0$ (not shown), photon fluctuations erase the entire NP (except when interactions are reciprocal, i.e., $\phi = 0,\frac{\pi}{2}$) \cite{Chiacchio2019b} so, strictly speaking, this case does not correspond to a NRPT; we nevertheless recover a NRPT for any small perturbation $\delta, \Gamma_{\downarrow}\neq0$.
%via incoherent processes. 
%which also competes with the formation of limit cycles. 
Second, and most strikingly, the NP and DP are no longer separated by a boundary of EPs. Instead, the fluctuations of the light field soften this feature, resulting in a smooth spectrum as a function of $\phi$~\cite{SuppMatt}, see Fig.~\ref{PD-Spec}(b). We insist that this still corresponds to a NRPT, as it is non-reciprocal interactions that give rise to the dynamical phase. We expect this behavior to emerge whenever non-reciprocity is mediated by dynamical degrees of freedom.

\begin{figure}[t]
\centering
\includegraphics[width=\columnwidth]{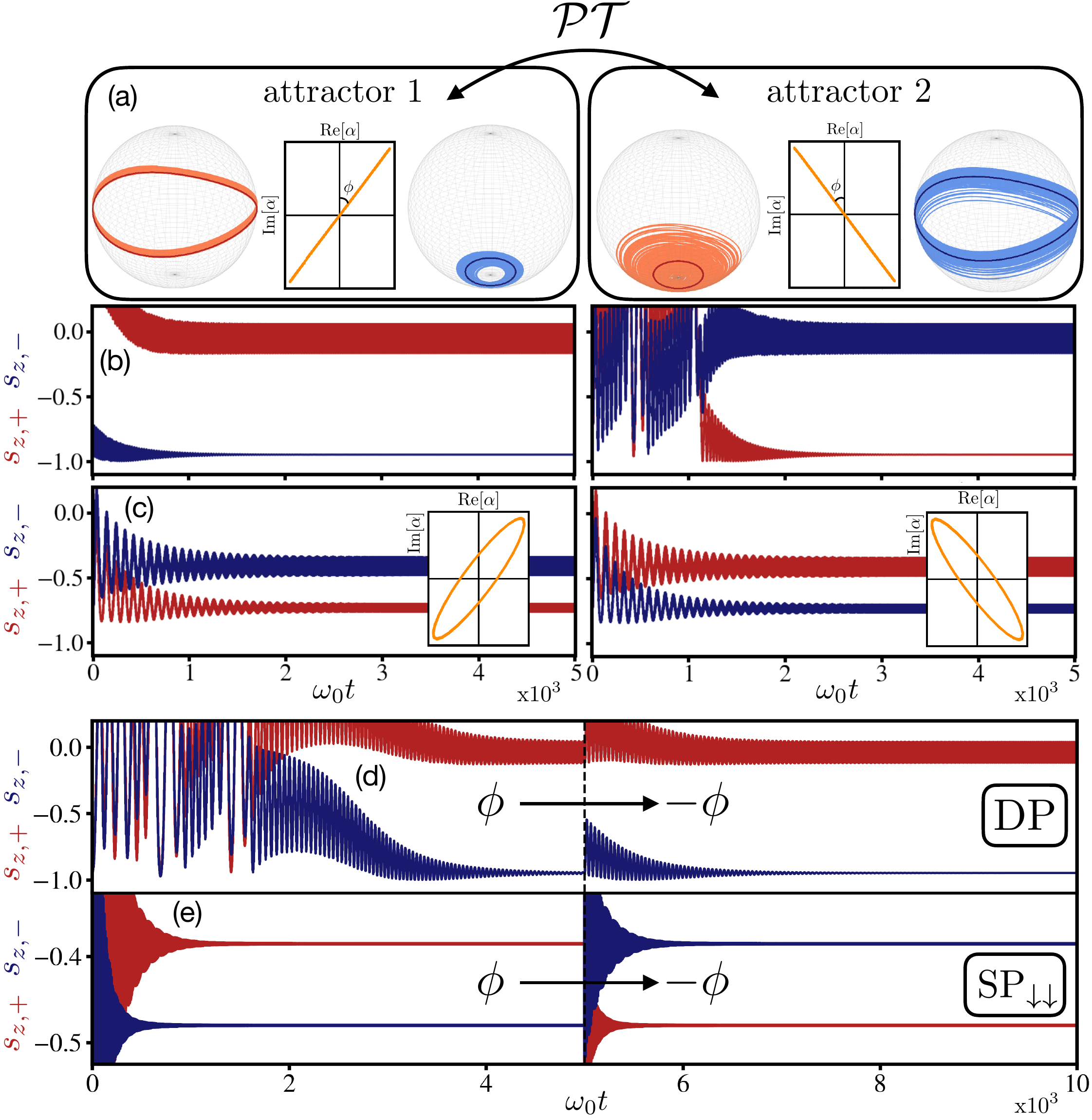}
\caption{\label{Time_Dynamics}  
\textbf{Dynamics and $\mathcal{PT}$ symmetry breaking.}
(a) The two steady-state attractors of Eqs.~\eqref{eq:er2} are related by a $\mathcal{PT}$ transformation, for $\phi = \pi/4$, $\lambda = 3\omega_{0}$ and $\delta = \Gamma_{\downarrow} = 0$; the light field phase locks at the angles $\frac{\pi}{2}\pm \phi$ and the spin trajectories of the two species are depicted on the Bloch sphere. The components $s_{z,\pm}$ are shown in (b) and in (c) for $\Gamma_{\downarrow} \neq 0$. (d)-(e) After the transient has elapsed, quenching the phase $\phi\rightarrow-\phi$ reveals that the DP is a $\mathcal{PT}$ broken phase (d) and SP$_{\downarrow\downarrow}$ is $\mathcal{PT}$ unbroken (e). Parameters: (a)-(e) $\omega_{l} = 20\omega_{0}$, $\kappa = 12.5\omega_{0}$; (d) $\lambda = 2.5\omega_{0}$, $\phi = \frac{\pi}{4}$; (e) $\lambda = 5.5 \omega_{0}$, $\phi = \frac{\pi}{8}$.}
\end{figure}

\begin{figure}[t]
\centering
\includegraphics[width=\columnwidth]{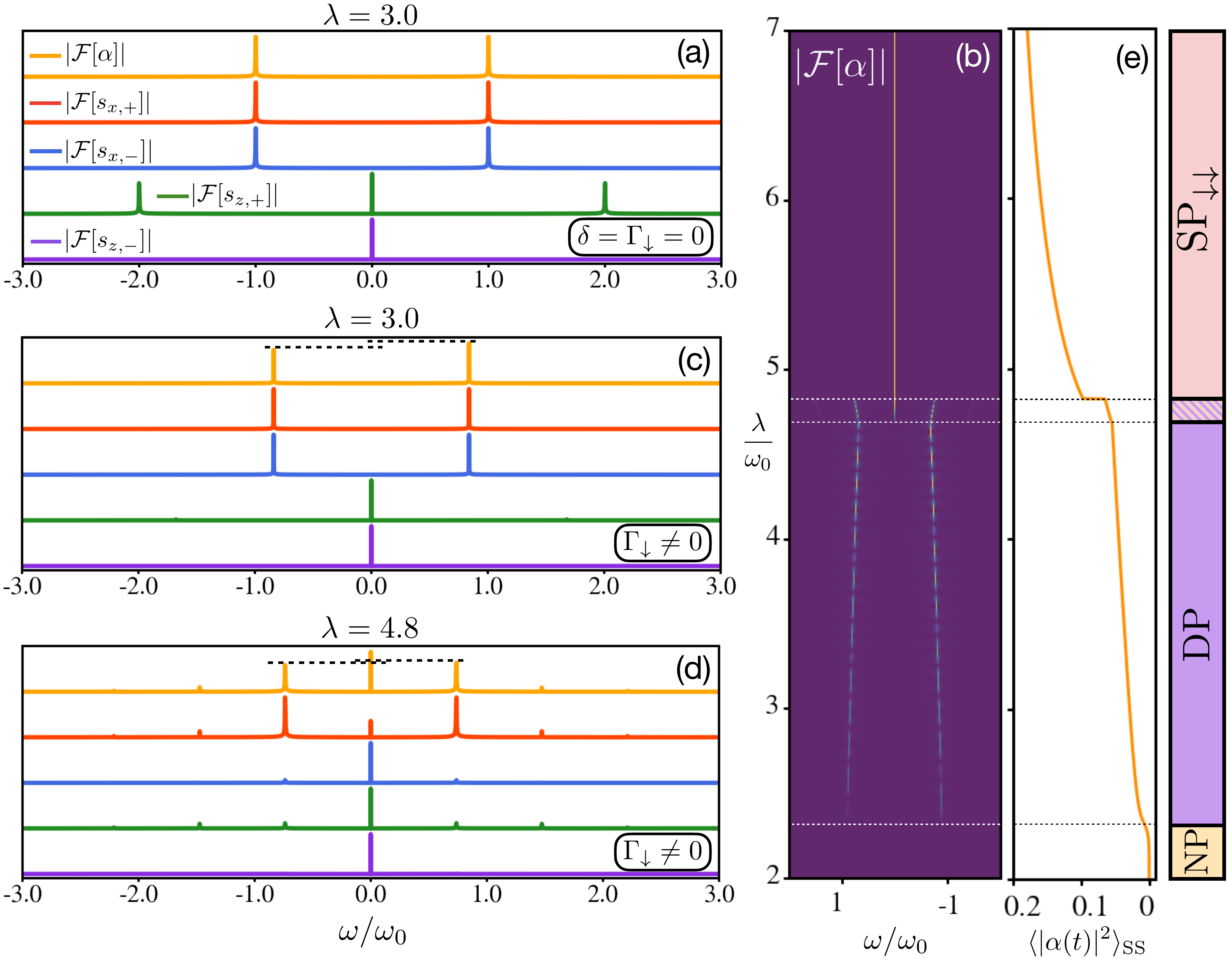}
\caption{\label{Spectrum} \textbf{Frequency spectrum}.
Absolute value of the Fourier spectrum $\mathcal{F}$ (arbitrary units) of the time evolution inside the DP for (a),(c),(d) the light field and $x$ and $z$ spin components, and (b) for the light field as a function of $\lambda$. The color shading in (b) is saturated at a cutoff value for better visualization.
Dashed lines in (c) and (d) highlight the height difference between peaks in the spectrum.
(e) Time-averaged intensity of the light field in the steady state as a function of $\lambda$. Parameters: (a)-(e) $\omega_{l}  = 20\omega_{0}$, $\kappa = 12.5\omega_{0}$ and $\phi = \pi/5$; (b)-(e) $\Gamma_{\downarrow}= 0.02\omega_{0}$.}
\end{figure}

\textit{Steady-state dynamics and $\mathcal{PT}$ symmetry breaking.---}
We now connect the occurrence of a NRPT to the spontaneous breaking of $\mathcal{PT}$ symmetry. Inside the DP, spins and light field undergo persistent oscillations in the form of limit cycles \cite{Chiacchio2019b}.
 
For $\delta= \Gamma_{\downarrow}=0$, the long-time dynamics is in fact determined by \emph{two} limit-cycle attractors, shown in Fig.~\ref{Time_Dynamics}(a); the $z$-component of the Bloch sphere trajectories is further highlighted in panel (b). We notice that light field oscillates with the phase locked at the angles $\frac{\pi}{2}\pm \phi$. Comparing the two attractor solutions we see that they are related by an exchange of the two species and a change of sign in the phase, i.e., via $\mathcal{PT}$. Depending on the initial conditions, the system settles into one of the two available steady states: the $\mathcal{PT}$ symmetry of Eqs.~\eqref{eq:er2} is therefore broken in the steady state, namely the onset of the DP is accompanied by the spontaneous breaking of $\mathcal{PT}$ symmetry. We stress that, due to the nonlinear character of Eqs.~\eqref{eq:er2}, this is truly a \emph{spontaneous} breaking of $\mathcal{PT}$ symmetry, unlike for $\mathcal{PT}$ symmetric linear systems.

Noticeably, for $\Gamma_{\downarrow} \neq 0$, $\delta=0$, Eqs.~\eqref{eq:er2} are still  $\mathcal{PT}$ symmetric, as spin decay acts homogeneously on the spins, yielding again two different steady-state attractors and a $\mathcal{PT}$ breaking phase transition. As shown in Fig.~\ref{Time_Dynamics}(c), both projections $s_{z,\pm}$ now display oscillations, while the light displays imperfect phase locking, however still about well defined angles $\frac{\pi}{2}\pm \phi$. We conclude that the NRPT present in the phase diagram Fig.~\ref{PD-Spec}(c) is accompanied by the spontaneous breaking of $\mathcal{PT}$ symmetry. In contrast, a finite frequency imbalance $\delta \neq 0$ explicitly breaks the $\mathcal{PT}$ symmetry of the NRDM at the level of the equations of motion, resulting in the erasure of one of the attractors, and leaving only a single non-stationary solution in the steady-state dynamics (not shown).

To show that $\mathcal{PT}$ symmetry breaking is a phenomenon uniquely associated to NRPTs, and not to normal-to-superradiant phase transitions, we perform additional simulations, in which we first let the system relax to the steady state then quench the phase $\phi \rightarrow -\phi$. When the system is in the DP, following the quench, it relaxes back to the same attractor, see Fig.~\ref{Time_Dynamics}(d). This characterizes the steady state as a $\mathcal{PT}$ broken state, since a further swap of the spin species (exchanging their colors) shows that the action of $\mathcal{PT}$ does not leave the steady state invariant. In panel (e) we perform the same numerical experiment starting from  SP$_{\downarrow\downarrow}$, from which we see that the action parity and time reversal (exchanging the species and reversing the sign of $\phi$) undo each other, i.e., the superradiant steady state is $\mathcal{PT}$ invariant; the same conclusion applies when starting from SP$_{\uparrow\downarrow}$ and NP.

\emph{Frequency spectrum and dynamical superradiance.}--- Another remarkable feature shown in Fig.~\ref{Time_Dynamics}(a), (b) is that, inside each attractor, the trajectories of the two spins are qualitatively different, with one species oscillating in-plane with constant $s_{z}$, and the other one featuring oscillations also along $s_{z}$. To obtain further insight, we perform a Fourier decomposition of the steady-state dynamics for the case $\delta = \Gamma_{\downarrow} = 0$, and we find the $(x,y)$ components of both species to oscillate at the natural frequency $\omega_{0}$, while the non-stationary $z$ component precesses at frequency $2\omega_{0}$, see Fig.~\ref{Spectrum}(a). This behavior can be understood as follows:
the phase of the light field locks at the angles $\frac{\pi}{2} \pm \phi$ and decouples from the spin species $\pm$ by spontaneously choosing either $\alpha e^{\mp i\phi} + \alpha^{*} e^{\pm i\phi} = 0$. This naturally leads to in-plane (constant $s_{z}$) coherent dynamics at $\omega_{0}$ only for the species $\pm$. We also find that the light field oscillation frequency locks to the natural frequency of the spins, as visible from the corresponding spectrum in Fig.~\ref{Spectrum}(a), which  peaks at $\pm \omega_{0}$~\cite{SuppMatt}.

The frequency analysis provides us with additional information and allows us to uncover new features of the NRDM, beyond those captured by the stability analysis. In Fig.~\ref{Spectrum}(b) we show the light frequency spectrum as a function of $\lambda$, for $\phi=\pi/5$, $\Gamma_{\downarrow} \neq 0$, $\delta=0$; the case  corresponds to the phase diagram in Fig.~\ref{PD-Spec}(d). Deep into the DP, i.e., for sufficiently small values of the coupling, we observe simple harmonic motion, with positive- and negative-frequency component that move closer to each other for increasing $\lambda$. The cut in Fig.~\ref{Spectrum}(c) further shows an asymmetry in the peaks of the two frequency components, in agreement with the ellipses in the inset of Fig.~\ref{Time_Dynamics}(c), and that the spin motion also locks at the same frequency. For large coupling values, on the other hand, we recover the static phase SP$_{\downarrow\downarrow}$, with its zero-frequency component. However, close to the SP we also find an intermediate range of values where the steady states simultaneously break both parity symmetry and $\mathcal{PT}$ symmetry: although competing, the two symmetry breaking processes are not mutually exclusive. Correspondingly, we observe the two main light frequency components repelling each other, see Fig.~\ref{Spectrum}(b), and the spin motion acquiring additional frequency components, see Fig.~\ref{Spectrum}(d). The coexistence of non-stationary and superradiant behavior, or \emph{dynamical superradiance} for short, is another remarkable finding of our work. The time-averaged intensity of the light field, shown in Fig.~\ref{Spectrum}(e), provides further information on the superradiant behavior of the NRDM: the transition to static superradiance (SP$_{\downarrow\downarrow}$) is marked by an abrupt jump, while the onset of dynamical superradiance occurs without jumps but is signalled by an increase in the steepness of the intensity as a function of $\lambda$, compared to the rest of the DP.

Finally, for values of the phase approaching $\phi= \pi/4$, i.e., maximum non-reciprocity, we also find that the regular attractors are lost and, as $\lambda$ increases, the system cascades into a chaotic regime with many emerging frequencies, which dominates until the superradiant phase transition occurs (not shown). An in-depth investigation of these regimes is left for future studies.

\textit{Conclusion.---}
We introduced the non-reciprocal Dicke model (NRDM) as a minimal setting to study non-reciprocal interactions in many-body light-matter systems. We identified the presence of a non-reciprocal phase transition (NRPT), showed that NRPTs can occur in a broader class of systems than previously known~\cite{Fruchart2021}, and linked the NRPT to spontaneous breaking of $\mathcal{PT}$ symmetry at the level of steady states.

Our results can be tested in state-of-the-art atom-cavity experiments \cite{Weiner2017, EsslingerSpin, Dogra2019, Kroeze2018, KeelingLevPRX, Periwal2021}.
These platforms further provide the opportunity to explore the effects of finite-range interactions \cite{KeelingLevPRX, Periwal2021}, or the possibility to include a lattice and probe the competition between non-reciprocal and Hubbard-type interactions \cite{Esslinger2}.
From a theory point of view, it would also be interesting to contrast non-reciprocal interactions with other ways to stabilise limit-cycle phases in Dicke-type models~\cite{Chitra2015,Zhu2019,Chelpanova2021}.

\begin{acknowledgments}
M.B.~acknowledges funding from the Swiss National Science Foundation under grant No.~PCEFP2\_194268.
\end{acknowledgments}

\bibliographystyle{apsrev4-2}
\bibliography{library}

\begin{thebibliography}{45}%
\makeatletter
\providecommand \@ifxundefined [1]{%
 \@ifx{#1\undefined}
}%
\providecommand \@ifnum [1]{%
 \ifnum #1\expandafter \@firstoftwo
 \else \expandafter \@secondoftwo
 \fi
}%
\providecommand \@ifx [1]{%
 \ifx #1\expandafter \@firstoftwo
 \else \expandafter \@secondoftwo
 \fi
}%
\providecommand \natexlab [1]{#1}%
\providecommand \enquote  [1]{``#1''}%
\providecommand \bibnamefont  [1]{#1}%
\providecommand \bibfnamefont [1]{#1}%
\providecommand \citenamefont [1]{#1}%
\providecommand \href@noop [0]{\@secondoftwo}%
\providecommand \href [0]{\begingroup \@sanitize@url \@href}%
\providecommand \@href[1]{\@@startlink{#1}\@@href}%
\providecommand \@@href[1]{\endgroup#1\@@endlink}%
\providecommand \@sanitize@url [0]{\catcode `\\12\catcode `\$12\catcode
  `\&12\catcode `\#12\catcode `\^12\catcode `\_12\catcode `\%12\relax}%
\providecommand \@@startlink[1]{}%
\providecommand \@@endlink[0]{}%
\providecommand \url  [0]{\begingroup\@sanitize@url \@url }%
\providecommand \@url [1]{\endgroup\@href {#1}{\urlprefix }}%
\providecommand \urlprefix  [0]{URL }%
\providecommand \Eprint [0]{\href }%
\providecommand \doibase [0]{https://doi.org/}%
\providecommand \selectlanguage [0]{\@gobble}%
\providecommand \bibinfo  [0]{\@secondoftwo}%
\providecommand \bibfield  [0]{\@secondoftwo}%
\providecommand \translation [1]{[#1]}%
\providecommand \BibitemOpen [0]{}%
\providecommand \bibitemStop [0]{}%
\providecommand \bibitemNoStop [0]{.\EOS\space}%
\providecommand \EOS [0]{\spacefactor3000\relax}%
\providecommand \BibitemShut  [1]{\csname bibitem#1\endcsname}%
\let\auto@bib@innerbib\@empty
%</preamble>
\bibitem [{\citenamefont {Newton}(1687)}]{Newton1687}%
  \BibitemOpen
  \bibfield  {author} {\bibinfo {author} {\bibfnamefont {I.}~\bibnamefont
  {Newton}},\ }\href@noop {} {\emph {\bibinfo {title} {Philosophiæ Naturalis
  Principia Mathematica}}}\ (\bibinfo {year} {1687})\BibitemShut {NoStop}%
\bibitem [{\citenamefont {Uchida}\ and\ \citenamefont
  {Golestanian}(2010)}]{Uchida2010}%
  \BibitemOpen
  \bibfield  {author} {\bibinfo {author} {\bibfnamefont {N.}~\bibnamefont
  {Uchida}}\ and\ \bibinfo {author} {\bibfnamefont {R.}~\bibnamefont
  {Golestanian}},\ }\href {https://doi.org/10.1103/PhysRevLett.104.178103}
  {\bibfield  {journal} {\bibinfo  {journal} {Physical Review Letters}\
  }\textbf {\bibinfo {volume} {104}},\ \bibinfo {pages} {178103} (\bibinfo
  {year} {2010})}\BibitemShut {NoStop}%
\bibitem [{\citenamefont {Nagy}\ \emph {et~al.}(2010)\citenamefont {Nagy},
  \citenamefont {Ákos}, \citenamefont {Biro},\ and\ \citenamefont
  {Vicsek}}]{Nagy2010}%
  \BibitemOpen
  \bibfield  {author} {\bibinfo {author} {\bibfnamefont {M.}~\bibnamefont
  {Nagy}}, \bibinfo {author} {\bibfnamefont {Z.}~\bibnamefont {Ákos}},
  \bibinfo {author} {\bibfnamefont {D.}~\bibnamefont {Biro}},\ and\ \bibinfo
  {author} {\bibfnamefont {T.}~\bibnamefont {Vicsek}},\ }\href
  {https://doi.org/10.1038/nature08891} {\bibfield  {journal} {\bibinfo
  {journal} {Nature}\ }\textbf {\bibinfo {volume} {464}},\ \bibinfo {pages}
  {890} (\bibinfo {year} {2010})}\BibitemShut {NoStop}%
\bibitem [{\citenamefont {Yllanes}\ \emph {et~al.}(2017)\citenamefont
  {Yllanes}, \citenamefont {Leoni},\ and\ \citenamefont
  {Marchetti}}]{Yllanes2017}%
  \BibitemOpen
  \bibfield  {author} {\bibinfo {author} {\bibfnamefont {D.}~\bibnamefont
  {Yllanes}}, \bibinfo {author} {\bibfnamefont {M.}~\bibnamefont {Leoni}},\
  and\ \bibinfo {author} {\bibfnamefont {M.~C.}\ \bibnamefont {Marchetti}},\
  }\href {https://doi.org/10.1088/1367-2630/aa8ed7} {\bibfield  {journal}
  {\bibinfo  {journal} {New Journal of Physics}\ }\textbf {\bibinfo {volume}
  {19}},\ \bibinfo {pages} {103026} (\bibinfo {year} {2017})}\BibitemShut
  {NoStop}%
\bibitem [{\citenamefont {Lavergne}\ \emph {et~al.}(2019)\citenamefont
  {Lavergne}, \citenamefont {Wendehenne}, \citenamefont {Bäuerle},\ and\
  \citenamefont {Bechinger}}]{Lavergne2019}%
  \BibitemOpen
  \bibfield  {author} {\bibinfo {author} {\bibfnamefont {F.~A.}\ \bibnamefont
  {Lavergne}}, \bibinfo {author} {\bibfnamefont {H.}~\bibnamefont
  {Wendehenne}}, \bibinfo {author} {\bibfnamefont {T.}~\bibnamefont
  {Bäuerle}},\ and\ \bibinfo {author} {\bibfnamefont {C.}~\bibnamefont
  {Bechinger}},\ }\href {https://doi.org/10.1126/science.aau5347} {\bibfield
  {journal} {\bibinfo  {journal} {Science}\ }\textbf {\bibinfo {volume}
  {364}},\ \bibinfo {pages} {70} (\bibinfo {year} {2019})}\BibitemShut
  {NoStop}%
\bibitem [{\citenamefont {Saha}\ \emph {et~al.}(2019)\citenamefont {Saha},
  \citenamefont {Ramaswamy},\ and\ \citenamefont {Golestanian}}]{Saha2019}%
  \BibitemOpen
  \bibfield  {author} {\bibinfo {author} {\bibfnamefont {S.}~\bibnamefont
  {Saha}}, \bibinfo {author} {\bibfnamefont {S.}~\bibnamefont {Ramaswamy}},\
  and\ \bibinfo {author} {\bibfnamefont {R.}~\bibnamefont {Golestanian}},\
  }\href {https://doi.org/10.1088/1367-2630/ab20fd} {\bibfield  {journal}
  {\bibinfo  {journal} {New Journal of Physics}\ }\textbf {\bibinfo {volume}
  {21}},\ \bibinfo {pages} {063006} (\bibinfo {year} {2019})}\BibitemShut
  {NoStop}%
\bibitem [{\citenamefont {Shankar}\ \emph {et~al.}(2022)\citenamefont
  {Shankar}, \citenamefont {Souslov}, \citenamefont {Bowick}, \citenamefont
  {Marchetti},\ and\ \citenamefont {Vitelli}}]{Shankar2022}%
  \BibitemOpen
  \bibfield  {author} {\bibinfo {author} {\bibfnamefont {S.}~\bibnamefont
  {Shankar}}, \bibinfo {author} {\bibfnamefont {A.}~\bibnamefont {Souslov}},
  \bibinfo {author} {\bibfnamefont {M.~J.}\ \bibnamefont {Bowick}}, \bibinfo
  {author} {\bibfnamefont {M.~C.}\ \bibnamefont {Marchetti}},\ and\ \bibinfo
  {author} {\bibfnamefont {V.}~\bibnamefont {Vitelli}},\ }\href
  {https://doi.org/10.1038/s42254-022-00445-3} {\bibfield  {journal} {\bibinfo
  {journal} {Nature Reviews Physics}\ }\textbf {\bibinfo {volume} {4}},\
  \bibinfo {pages} {380} (\bibinfo {year} {2022})}\BibitemShut {NoStop}%
\bibitem [{\citenamefont {Baconnier}\ \emph {et~al.}(2022)\citenamefont
  {Baconnier}, \citenamefont {Shohat}, \citenamefont {L{\'o}pez}, \citenamefont
  {Coulais}, \citenamefont {D{\'e}mery}, \citenamefont {D{\"u}ring},\ and\
  \citenamefont {Dauchot}}]{Baconnier2022}%
  \BibitemOpen
  \bibfield  {author} {\bibinfo {author} {\bibfnamefont {P.}~\bibnamefont
  {Baconnier}}, \bibinfo {author} {\bibfnamefont {D.}~\bibnamefont {Shohat}},
  \bibinfo {author} {\bibfnamefont {C.~H.}\ \bibnamefont {L{\'o}pez}}, \bibinfo
  {author} {\bibfnamefont {C.}~\bibnamefont {Coulais}}, \bibinfo {author}
  {\bibfnamefont {V.}~\bibnamefont {D{\'e}mery}}, \bibinfo {author}
  {\bibfnamefont {G.}~\bibnamefont {D{\"u}ring}},\ and\ \bibinfo {author}
  {\bibfnamefont {O.}~\bibnamefont {Dauchot}},\ }\href
  {https://doi.org/10.1038/s41567-022-01704-x} {\bibfield  {journal} {\bibinfo
  {journal} {Nature Physics}\ }\textbf {\bibinfo {volume} {18}},\ \bibinfo
  {pages} {1234} (\bibinfo {year} {2022})}\BibitemShut {NoStop}%
\bibitem [{\citenamefont {Coullet}\ \emph {et~al.}(1989)\citenamefont
  {Coullet}, \citenamefont {Goldstein},\ and\ \citenamefont
  {Gunaratne}}]{Coullet1989}%
  \BibitemOpen
  \bibfield  {author} {\bibinfo {author} {\bibfnamefont {P.}~\bibnamefont
  {Coullet}}, \bibinfo {author} {\bibfnamefont {R.~E.}\ \bibnamefont
  {Goldstein}},\ and\ \bibinfo {author} {\bibfnamefont {G.~H.}\ \bibnamefont
  {Gunaratne}},\ }\href {https://doi.org/10.1103/PhysRevLett.63.1954}
  {\bibfield  {journal} {\bibinfo  {journal} {Physical Review Letters}\
  }\textbf {\bibinfo {volume} {63}},\ \bibinfo {pages} {1954} (\bibinfo {year}
  {1989})}\BibitemShut {NoStop}%
\bibitem [{\citenamefont {Pan}\ and\ \citenamefont {de~Bruyn}(1994)}]{Pan1994}%
  \BibitemOpen
  \bibfield  {author} {\bibinfo {author} {\bibfnamefont {L.}~\bibnamefont
  {Pan}}\ and\ \bibinfo {author} {\bibfnamefont {J.~R.}\ \bibnamefont
  {de~Bruyn}},\ }\href {https://doi.org/10.1103/PhysRevE.49.483} {\bibfield
  {journal} {\bibinfo  {journal} {Physical Review E}\ }\textbf {\bibinfo
  {volume} {49}},\ \bibinfo {pages} {483} (\bibinfo {year} {1994})}\BibitemShut
  {NoStop}%
\bibitem [{\citenamefont {Montbrió}\ and\ \citenamefont
  {Pazó}(2018)}]{Montbrio2018}%
  \BibitemOpen
  \bibfield  {author} {\bibinfo {author} {\bibfnamefont {E.}~\bibnamefont
  {Montbrió}}\ and\ \bibinfo {author} {\bibfnamefont {D.}~\bibnamefont
  {Pazó}},\ }\href {https://doi.org/10.1103/PhysRevLett.120.244101} {\bibfield
   {journal} {\bibinfo  {journal} {Physical Review Letters}\ }\textbf {\bibinfo
  {volume} {120}},\ \bibinfo {pages} {244101} (\bibinfo {year}
  {2018})}\BibitemShut {NoStop}%
\bibitem [{\citenamefont {Hong}\ and\ \citenamefont
  {Strogatz}(2011)}]{Hong2011}%
  \BibitemOpen
  \bibfield  {author} {\bibinfo {author} {\bibfnamefont {H.}~\bibnamefont
  {Hong}}\ and\ \bibinfo {author} {\bibfnamefont {S.~H.}\ \bibnamefont
  {Strogatz}},\ }\href {https://doi.org/10.1103/PhysRevLett.106.054102}
  {\bibfield  {journal} {\bibinfo  {journal} {Physical Review Letters}\
  }\textbf {\bibinfo {volume} {106}},\ \bibinfo {pages} {054102} (\bibinfo
  {year} {2011})}\BibitemShut {NoStop}%
\bibitem [{\citenamefont {Fruchart}\ \emph {et~al.}(2021)\citenamefont
  {Fruchart}, \citenamefont {Hanai}, \citenamefont {Littlewood},\ and\
  \citenamefont {Vitelli}}]{Fruchart2021}%
  \BibitemOpen
  \bibfield  {author} {\bibinfo {author} {\bibfnamefont {M.}~\bibnamefont
  {Fruchart}}, \bibinfo {author} {\bibfnamefont {R.}~\bibnamefont {Hanai}},
  \bibinfo {author} {\bibfnamefont {P.~B.}\ \bibnamefont {Littlewood}},\ and\
  \bibinfo {author} {\bibfnamefont {V.}~\bibnamefont {Vitelli}},\ }\href
  {https://doi.org/10.1038/s41586-021-03375-9} {\bibfield  {journal} {\bibinfo
  {journal} {Nature}\ }\textbf {\bibinfo {volume} {592}},\ \bibinfo {pages}
  {363} (\bibinfo {year} {2021})}\BibitemShut {NoStop}%
\bibitem [{\citenamefont {Jalas}\ \emph {et~al.}(2013)\citenamefont {Jalas},
  \citenamefont {Petrov}, \citenamefont {Eich}, \citenamefont {Freude},
  \citenamefont {Fan}, \citenamefont {Yu}, \citenamefont {Baets}, \citenamefont
  {Popović}, \citenamefont {Melloni}, \citenamefont {Joannopoulos},
  \citenamefont {Vanwolleghem}, \citenamefont {Doerr},\ and\ \citenamefont
  {Renner}}]{Jalas2013}%
  \BibitemOpen
  \bibfield  {author} {\bibinfo {author} {\bibfnamefont {D.}~\bibnamefont
  {Jalas}}, \bibinfo {author} {\bibfnamefont {A.}~\bibnamefont {Petrov}},
  \bibinfo {author} {\bibfnamefont {M.}~\bibnamefont {Eich}}, \bibinfo {author}
  {\bibfnamefont {W.}~\bibnamefont {Freude}}, \bibinfo {author} {\bibfnamefont
  {S.}~\bibnamefont {Fan}}, \bibinfo {author} {\bibfnamefont {Z.}~\bibnamefont
  {Yu}}, \bibinfo {author} {\bibfnamefont {R.}~\bibnamefont {Baets}}, \bibinfo
  {author} {\bibfnamefont {M.}~\bibnamefont {Popović}}, \bibinfo {author}
  {\bibfnamefont {A.}~\bibnamefont {Melloni}}, \bibinfo {author} {\bibfnamefont
  {J.~D.}\ \bibnamefont {Joannopoulos}}, \bibinfo {author} {\bibfnamefont
  {M.}~\bibnamefont {Vanwolleghem}}, \bibinfo {author} {\bibfnamefont {C.~R.}\
  \bibnamefont {Doerr}},\ and\ \bibinfo {author} {\bibfnamefont
  {H.}~\bibnamefont {Renner}},\ }\href
  {https://doi.org/10.1038/nphoton.2013.185 http://10.0.4.14/nphoton.2013.185}
  {\bibfield  {journal} {\bibinfo  {journal} {Nature Photonics}\ }\textbf
  {\bibinfo {volume} {7}},\ \bibinfo {pages} {579} (\bibinfo {year}
  {2013})}\BibitemShut {NoStop}%
\bibitem [{\citenamefont {Fleury}\ \emph {et~al.}(2014)\citenamefont {Fleury},
  \citenamefont {Sounas}, \citenamefont {Sieck}, \citenamefont {Haberman},\
  and\ \citenamefont {Alù}}]{Fleury2014}%
  \BibitemOpen
  \bibfield  {author} {\bibinfo {author} {\bibfnamefont {R.}~\bibnamefont
  {Fleury}}, \bibinfo {author} {\bibfnamefont {D.~L.}\ \bibnamefont {Sounas}},
  \bibinfo {author} {\bibfnamefont {C.~F.}\ \bibnamefont {Sieck}}, \bibinfo
  {author} {\bibfnamefont {M.~R.}\ \bibnamefont {Haberman}},\ and\ \bibinfo
  {author} {\bibfnamefont {A.}~\bibnamefont {Alù}},\ }\href
  {https://doi.org/10.1126/science.1246957} {\bibfield  {journal} {\bibinfo
  {journal} {Science}\ }\textbf {\bibinfo {volume} {343}},\ \bibinfo {pages}
  {516} (\bibinfo {year} {2014})}\BibitemShut {NoStop}%
\bibitem [{\citenamefont {Estep}\ \emph {et~al.}(2014)\citenamefont {Estep},
  \citenamefont {Sounas}, \citenamefont {Soric},\ and\ \citenamefont
  {Alù}}]{Estep2014}%
  \BibitemOpen
  \bibfield  {author} {\bibinfo {author} {\bibfnamefont {N.~A.}\ \bibnamefont
  {Estep}}, \bibinfo {author} {\bibfnamefont {D.~L.}\ \bibnamefont {Sounas}},
  \bibinfo {author} {\bibfnamefont {J.}~\bibnamefont {Soric}},\ and\ \bibinfo
  {author} {\bibfnamefont {A.}~\bibnamefont {Alù}},\ }\href
  {https://doi.org/10.1038/nphys3134} {\bibfield  {journal} {\bibinfo
  {journal} {Nature Physics}\ }\textbf {\bibinfo {volume} {10}},\ \bibinfo
  {pages} {923} (\bibinfo {year} {2014})}\BibitemShut {NoStop}%
\bibitem [{\citenamefont {Metelmann}\ and\ \citenamefont
  {Clerk}(2015)}]{Metelmann2015}%
  \BibitemOpen
  \bibfield  {author} {\bibinfo {author} {\bibfnamefont {A.}~\bibnamefont
  {Metelmann}}\ and\ \bibinfo {author} {\bibfnamefont {A.}~\bibnamefont
  {Clerk}},\ }\href {https://doi.org/10.1103/PhysRevX.5.021025} {\bibfield
  {journal} {\bibinfo  {journal} {Physical Review X}\ }\textbf {\bibinfo
  {volume} {5}},\ \bibinfo {pages} {021025} (\bibinfo {year}
  {2015})}\BibitemShut {NoStop}%
\bibitem [{\citenamefont {Lodahl}\ \emph {et~al.}(2017)\citenamefont {Lodahl},
  \citenamefont {Mahmoodian}, \citenamefont {Stobbe}, \citenamefont
  {Rauschenbeutel}, \citenamefont {Schneeweiss}, \citenamefont {Volz},
  \citenamefont {Pichler},\ and\ \citenamefont {Zoller}}]{Lodahl2017}%
  \BibitemOpen
  \bibfield  {author} {\bibinfo {author} {\bibfnamefont {P.}~\bibnamefont
  {Lodahl}}, \bibinfo {author} {\bibfnamefont {S.}~\bibnamefont {Mahmoodian}},
  \bibinfo {author} {\bibfnamefont {S.}~\bibnamefont {Stobbe}}, \bibinfo
  {author} {\bibfnamefont {A.}~\bibnamefont {Rauschenbeutel}}, \bibinfo
  {author} {\bibfnamefont {P.}~\bibnamefont {Schneeweiss}}, \bibinfo {author}
  {\bibfnamefont {J.}~\bibnamefont {Volz}}, \bibinfo {author} {\bibfnamefont
  {H.}~\bibnamefont {Pichler}},\ and\ \bibinfo {author} {\bibfnamefont
  {P.}~\bibnamefont {Zoller}},\ }\href {https://doi.org/10.1038/nature21037
  http://10.0.4.14/nature21037} {\bibfield  {journal} {\bibinfo  {journal}
  {Nature}\ }\textbf {\bibinfo {volume} {541}},\ \bibinfo {pages} {473}
  (\bibinfo {year} {2017})}\BibitemShut {NoStop}%
\bibitem [{\citenamefont {Bernier}\ \emph {et~al.}(2017)\citenamefont
  {Bernier}, \citenamefont {Tóth}, \citenamefont {Koottandavida},
  \citenamefont {Ioannou}, \citenamefont {Malz}, \citenamefont {Nunnenkamp},
  \citenamefont {Feofanov},\ and\ \citenamefont {Kippenberg}}]{Bernier2017}%
  \BibitemOpen
  \bibfield  {author} {\bibinfo {author} {\bibfnamefont {N.~R.}\ \bibnamefont
  {Bernier}}, \bibinfo {author} {\bibfnamefont {L.~D.}\ \bibnamefont {Tóth}},
  \bibinfo {author} {\bibfnamefont {A.}~\bibnamefont {Koottandavida}}, \bibinfo
  {author} {\bibfnamefont {M.~A.}\ \bibnamefont {Ioannou}}, \bibinfo {author}
  {\bibfnamefont {D.}~\bibnamefont {Malz}}, \bibinfo {author} {\bibfnamefont
  {A.}~\bibnamefont {Nunnenkamp}}, \bibinfo {author} {\bibfnamefont {A.~K.}\
  \bibnamefont {Feofanov}},\ and\ \bibinfo {author} {\bibfnamefont {T.~J.}\
  \bibnamefont {Kippenberg}},\ }\href
  {https://doi.org/10.1038/s41467-017-00447-1} {\bibfield  {journal} {\bibinfo
  {journal} {Nature Communications}\ }\textbf {\bibinfo {volume} {8}},\
  \bibinfo {pages} {604} (\bibinfo {year} {2017})}\BibitemShut {NoStop}%
\bibitem [{\citenamefont {Verhagen}\ and\ \citenamefont
  {Alù}(2017)}]{Verhagen2017}%
  \BibitemOpen
  \bibfield  {author} {\bibinfo {author} {\bibfnamefont {E.}~\bibnamefont
  {Verhagen}}\ and\ \bibinfo {author} {\bibfnamefont {A.}~\bibnamefont
  {Alù}},\ }\href {https://doi.org/10.1038/nphys4283
  http://10.0.4.14/nphys4283} {\bibfield  {journal} {\bibinfo  {journal}
  {Nature Physics}\ }\textbf {\bibinfo {volume} {13}},\ \bibinfo {pages} {922}
  (\bibinfo {year} {2017})}\BibitemShut {NoStop}%
\bibitem [{\citenamefont {Miri}\ and\ \citenamefont {Alù}(2019)}]{Miri2019}%
  \BibitemOpen
  \bibfield  {author} {\bibinfo {author} {\bibfnamefont {M.-A.}\ \bibnamefont
  {Miri}}\ and\ \bibinfo {author} {\bibfnamefont {A.}~\bibnamefont {Alù}},\
  }\href {https://doi.org/10.1126/science.aar7709} {\bibfield  {journal}
  {\bibinfo  {journal} {Science}\ }\textbf {\bibinfo {volume} {363}},\ \bibinfo
  {pages} {eaar7709} (\bibinfo {year} {2019})}\BibitemShut {NoStop}%
\bibitem [{\citenamefont {Hanai}\ \emph {et~al.}(2019)\citenamefont {Hanai},
  \citenamefont {Edelman}, \citenamefont {Ohashi},\ and\ \citenamefont
  {Littlewood}}]{HanaiPRL2019}%
  \BibitemOpen
  \bibfield  {author} {\bibinfo {author} {\bibfnamefont {R.}~\bibnamefont
  {Hanai}}, \bibinfo {author} {\bibfnamefont {A.}~\bibnamefont {Edelman}},
  \bibinfo {author} {\bibfnamefont {Y.}~\bibnamefont {Ohashi}},\ and\ \bibinfo
  {author} {\bibfnamefont {P.~B.}\ \bibnamefont {Littlewood}},\ }\href
  {https://doi.org/10.1103/PhysRevLett.122.185301} {\bibfield  {journal}
  {\bibinfo  {journal} {Physical Review Letters}\ }\textbf {\bibinfo {volume}
  {122}},\ \bibinfo {pages} {185301} (\bibinfo {year} {2019})}\BibitemShut
  {NoStop}%
\bibitem [{\citenamefont {Kirton}\ \emph {et~al.}(2019)\citenamefont {Kirton},
  \citenamefont {Roses}, \citenamefont {Keeling},\ and\ \citenamefont
  {Torre}}]{KeelingDicke}%
  \BibitemOpen
  \bibfield  {author} {\bibinfo {author} {\bibfnamefont {P.}~\bibnamefont
  {Kirton}}, \bibinfo {author} {\bibfnamefont {M.~M.}\ \bibnamefont {Roses}},
  \bibinfo {author} {\bibfnamefont {J.}~\bibnamefont {Keeling}},\ and\ \bibinfo
  {author} {\bibfnamefont {E.~G.~D.}\ \bibnamefont {Torre}},\ }\href
  {https://doi.org/10.1002/qute.201800043} {\bibfield  {journal} {\bibinfo
  {journal} {Advanced Quantum Technologies}\ }\textbf {\bibinfo {volume} {2}},\
  \bibinfo {pages} {1800043} (\bibinfo {year} {2019})}\BibitemShut {NoStop}%
\bibitem [{\citenamefont {Bender}(2007)}]{Bender2007}%
  \BibitemOpen
  \bibfield  {author} {\bibinfo {author} {\bibfnamefont {C.~M.}\ \bibnamefont
  {Bender}},\ }\href {https://doi.org/10.1088/0034-4885/70/6/R03} {\bibfield
  {journal} {\bibinfo  {journal} {Reports on Progress in Physics}\ }\textbf
  {\bibinfo {volume} {70}},\ \bibinfo {pages} {947} (\bibinfo {year}
  {2007})}\BibitemShut {NoStop}%
\bibitem [{\citenamefont {Chiacchio}\ and\ \citenamefont
  {Nunnenkamp}(2019)}]{Chiacchio2019b}%
  \BibitemOpen
  \bibfield  {author} {\bibinfo {author} {\bibfnamefont {E.~I.~R.}\
  \bibnamefont {Chiacchio}}\ and\ \bibinfo {author} {\bibfnamefont
  {A.}~\bibnamefont {Nunnenkamp}},\ }\href
  {https://doi.org/10.1103/PhysRevLett.122.193605} {\bibfield  {journal}
  {\bibinfo  {journal} {Physical Review Letters}\ }\textbf {\bibinfo {volume}
  {122}},\ \bibinfo {pages} {193605} (\bibinfo {year} {2019})}\BibitemShut
  {NoStop}%
\bibitem [{\citenamefont {Buča}\ and\ \citenamefont
  {Jaksch}(2019)}]{Buca2019}%
  \BibitemOpen
  \bibfield  {author} {\bibinfo {author} {\bibfnamefont {B.}~\bibnamefont
  {Buča}}\ and\ \bibinfo {author} {\bibfnamefont {D.}~\bibnamefont {Jaksch}},\
  }\href {https://doi.org/10.1103/PhysRevLett.123.260401} {\bibfield  {journal}
  {\bibinfo  {journal} {Physical Review Letters}\ }\textbf {\bibinfo {volume}
  {123}},\ \bibinfo {pages} {260401} (\bibinfo {year} {2019})}\BibitemShut
  {NoStop}%
\bibitem [{\citenamefont {Dogra}\ \emph {et~al.}(2019)\citenamefont {Dogra},
  \citenamefont {Landini}, \citenamefont {Kroeger}, \citenamefont {Hruby},
  \citenamefont {Donner},\ and\ \citenamefont {Esslinger}}]{Dogra2019}%
  \BibitemOpen
  \bibfield  {author} {\bibinfo {author} {\bibfnamefont {N.}~\bibnamefont
  {Dogra}}, \bibinfo {author} {\bibfnamefont {M.}~\bibnamefont {Landini}},
  \bibinfo {author} {\bibfnamefont {K.}~\bibnamefont {Kroeger}}, \bibinfo
  {author} {\bibfnamefont {L.}~\bibnamefont {Hruby}}, \bibinfo {author}
  {\bibfnamefont {T.}~\bibnamefont {Donner}},\ and\ \bibinfo {author}
  {\bibfnamefont {T.}~\bibnamefont {Esslinger}},\ }\href
  {https://doi.org/10.1126/science.aaw4465} {\bibfield  {journal} {\bibinfo
  {journal} {Science}\ }\textbf {\bibinfo {volume} {366}},\ \bibinfo {pages}
  {1496} (\bibinfo {year} {2019})}\BibitemShut {NoStop}%
\bibitem [{\citenamefont {Weiner}\ \emph {et~al.}(2017)\citenamefont {Weiner},
  \citenamefont {Cox}, \citenamefont {Bohnet},\ and\ \citenamefont
  {Thompson}}]{Weiner2017}%
  \BibitemOpen
  \bibfield  {author} {\bibinfo {author} {\bibfnamefont {J.~M.}\ \bibnamefont
  {Weiner}}, \bibinfo {author} {\bibfnamefont {K.~C.}\ \bibnamefont {Cox}},
  \bibinfo {author} {\bibfnamefont {J.~G.}\ \bibnamefont {Bohnet}},\ and\
  \bibinfo {author} {\bibfnamefont {J.~K.}\ \bibnamefont {Thompson}},\ }\href
  {https://doi.org/10.1103/PhysRevA.95.033808} {\bibfield  {journal} {\bibinfo
  {journal} {Physical Review A}\ }\textbf {\bibinfo {volume} {95}},\ \bibinfo
  {pages} {033808} (\bibinfo {year} {2017})}\BibitemShut {NoStop}%
\bibitem [{\citenamefont {Landini}\ \emph {et~al.}(2018)\citenamefont
  {Landini}, \citenamefont {Dogra}, \citenamefont {Kroeger}, \citenamefont
  {Hruby}, \citenamefont {Donner},\ and\ \citenamefont
  {Esslinger}}]{EsslingerSpin}%
  \BibitemOpen
  \bibfield  {author} {\bibinfo {author} {\bibfnamefont {M.}~\bibnamefont
  {Landini}}, \bibinfo {author} {\bibfnamefont {N.}~\bibnamefont {Dogra}},
  \bibinfo {author} {\bibfnamefont {K.}~\bibnamefont {Kroeger}}, \bibinfo
  {author} {\bibfnamefont {L.}~\bibnamefont {Hruby}}, \bibinfo {author}
  {\bibfnamefont {T.}~\bibnamefont {Donner}},\ and\ \bibinfo {author}
  {\bibfnamefont {T.}~\bibnamefont {Esslinger}},\ }\href
  {https://doi.org/10.1103/PhysRevLett.120.223602} {\bibfield  {journal}
  {\bibinfo  {journal} {Physical Review Letters}\ }\textbf {\bibinfo {volume}
  {120}},\ \bibinfo {pages} {223602} (\bibinfo {year} {2018})}\BibitemShut
  {NoStop}%
\bibitem [{\citenamefont {Kroeze}\ \emph {et~al.}(2018)\citenamefont {Kroeze},
  \citenamefont {Guo}, \citenamefont {Vaidya}, \citenamefont {Keeling},\ and\
  \citenamefont {Lev}}]{Kroeze2018}%
  \BibitemOpen
  \bibfield  {author} {\bibinfo {author} {\bibfnamefont {R.~M.}\ \bibnamefont
  {Kroeze}}, \bibinfo {author} {\bibfnamefont {Y.}~\bibnamefont {Guo}},
  \bibinfo {author} {\bibfnamefont {V.~D.}\ \bibnamefont {Vaidya}}, \bibinfo
  {author} {\bibfnamefont {J.}~\bibnamefont {Keeling}},\ and\ \bibinfo {author}
  {\bibfnamefont {B.~L.}\ \bibnamefont {Lev}},\ }\href
  {https://doi.org/10.1103/PhysRevLett.121.163601} {\bibfield  {journal}
  {\bibinfo  {journal} {Physical Review Letters}\ }\textbf {\bibinfo {volume}
  {121}},\ \bibinfo {pages} {163601} (\bibinfo {year} {2018})}\BibitemShut
  {NoStop}%
\bibitem [{\citenamefont {Vaidya}\ \emph {et~al.}(2018)\citenamefont {Vaidya},
  \citenamefont {Guo}, \citenamefont {Kroeze}, \citenamefont {Ballantine},
  \citenamefont {Kollár}, \citenamefont {Keeling},\ and\ \citenamefont
  {Lev}}]{KeelingLevPRX}%
  \BibitemOpen
  \bibfield  {author} {\bibinfo {author} {\bibfnamefont {V.~D.}\ \bibnamefont
  {Vaidya}}, \bibinfo {author} {\bibfnamefont {Y.}~\bibnamefont {Guo}},
  \bibinfo {author} {\bibfnamefont {R.~M.}\ \bibnamefont {Kroeze}}, \bibinfo
  {author} {\bibfnamefont {K.~E.}\ \bibnamefont {Ballantine}}, \bibinfo
  {author} {\bibfnamefont {A.~J.}\ \bibnamefont {Kollár}}, \bibinfo {author}
  {\bibfnamefont {J.}~\bibnamefont {Keeling}},\ and\ \bibinfo {author}
  {\bibfnamefont {B.~L.}\ \bibnamefont {Lev}},\ }\href
  {https://doi.org/10.1103/PhysRevX.8.011002} {\bibfield  {journal} {\bibinfo
  {journal} {Physical Review X}\ }\textbf {\bibinfo {volume} {8}},\ \bibinfo
  {pages} {011002} (\bibinfo {year} {2018})}\BibitemShut {NoStop}%
\bibitem [{\citenamefont {Periwal}\ \emph {et~al.}(2021)\citenamefont
  {Periwal}, \citenamefont {Cooper}, \citenamefont {Kunkel}, \citenamefont
  {Wienand}, \citenamefont {Davis},\ and\ \citenamefont
  {Schleier-Smith}}]{Periwal2021}%
  \BibitemOpen
  \bibfield  {author} {\bibinfo {author} {\bibfnamefont {A.}~\bibnamefont
  {Periwal}}, \bibinfo {author} {\bibfnamefont {E.~S.}\ \bibnamefont {Cooper}},
  \bibinfo {author} {\bibfnamefont {P.}~\bibnamefont {Kunkel}}, \bibinfo
  {author} {\bibfnamefont {J.~F.}\ \bibnamefont {Wienand}}, \bibinfo {author}
  {\bibfnamefont {E.~J.}\ \bibnamefont {Davis}},\ and\ \bibinfo {author}
  {\bibfnamefont {M.}~\bibnamefont {Schleier-Smith}},\ }\href
  {https://doi.org/10.1038/s41586-021-04156-0} {\bibfield  {journal} {\bibinfo
  {journal} {Nature}\ }\textbf {\bibinfo {volume} {600}},\ \bibinfo {pages}
  {630} (\bibinfo {year} {2021})}\BibitemShut {NoStop}%
\bibitem [{\citenamefont {Mivehvar}\ \emph {et~al.}(2021)\citenamefont
  {Mivehvar}, \citenamefont {Piazza}, \citenamefont {Donner},\ and\
  \citenamefont {Ritsch}}]{Mivehvar2021}%
  \BibitemOpen
  \bibfield  {author} {\bibinfo {author} {\bibfnamefont {F.}~\bibnamefont
  {Mivehvar}}, \bibinfo {author} {\bibfnamefont {F.}~\bibnamefont {Piazza}},
  \bibinfo {author} {\bibfnamefont {T.}~\bibnamefont {Donner}},\ and\ \bibinfo
  {author} {\bibfnamefont {H.}~\bibnamefont {Ritsch}},\ }\href
  {https://doi.org/10.1080/00018732.2021.1969727} {\bibfield  {journal}
  {\bibinfo  {journal} {Advances in Physics}\ }\textbf {\bibinfo {volume}
  {70}},\ \bibinfo {pages} {1} (\bibinfo {year} {2021})}\BibitemShut {NoStop}%
\bibitem [{\citenamefont {Hepp}\ and\ \citenamefont {Lieb}(1973)}]{HeppLieb}%
  \BibitemOpen
  \bibfield  {author} {\bibinfo {author} {\bibfnamefont {K.}~\bibnamefont
  {Hepp}}\ and\ \bibinfo {author} {\bibfnamefont {E.~H.}\ \bibnamefont
  {Lieb}},\ }\href {https://doi.org/10.1103/PhysRevA.8.2517} {\bibfield
  {journal} {\bibinfo  {journal} {Physical Review A}\ }\textbf {\bibinfo
  {volume} {8}},\ \bibinfo {pages} {2517} (\bibinfo {year} {1973})}\BibitemShut
  {NoStop}%
\bibitem [{\citenamefont {Wang}\ and\ \citenamefont {Hioe}(1973)}]{WangHioe}%
  \BibitemOpen
  \bibfield  {author} {\bibinfo {author} {\bibfnamefont {Y.~K.}\ \bibnamefont
  {Wang}}\ and\ \bibinfo {author} {\bibfnamefont {F.~T.}\ \bibnamefont
  {Hioe}},\ }\href {https://doi.org/10.1103/PhysRevA.7.831} {\bibfield
  {journal} {\bibinfo  {journal} {Physical Review A}\ }\textbf {\bibinfo
  {volume} {7}},\ \bibinfo {pages} {831} (\bibinfo {year} {1973})}\BibitemShut
  {NoStop}%
\bibitem [{Sup()}]{SuppMatt}%
  \BibitemOpen
  \href@noop {} {\bibinfo {title} {See supplemental material for additional
  discussion.}}\BibitemShut {Stop}%
\bibitem [{\citenamefont {Kepesidis}\ \emph {et~al.}(2016)\citenamefont
  {Kepesidis}, \citenamefont {Milburn}, \citenamefont {Huber}, \citenamefont
  {Makris}, \citenamefont {Rotter},\ and\ \citenamefont
  {Rabl}}]{Kepesidis2016}%
  \BibitemOpen
  \bibfield  {author} {\bibinfo {author} {\bibfnamefont {K.~V.}\ \bibnamefont
  {Kepesidis}}, \bibinfo {author} {\bibfnamefont {T.~J.}\ \bibnamefont
  {Milburn}}, \bibinfo {author} {\bibfnamefont {J.}~\bibnamefont {Huber}},
  \bibinfo {author} {\bibfnamefont {K.~G.}\ \bibnamefont {Makris}}, \bibinfo
  {author} {\bibfnamefont {S.}~\bibnamefont {Rotter}},\ and\ \bibinfo {author}
  {\bibfnamefont {P.}~\bibnamefont {Rabl}},\ }\href
  {https://doi.org/10.1088/1367-2630/18/9/095003} {\bibfield  {journal}
  {\bibinfo  {journal} {New Journal of Physics}\ }\textbf {\bibinfo {volume}
  {18}},\ \bibinfo {pages} {095003} (\bibinfo {year} {2016})}\BibitemShut
  {NoStop}%
\bibitem [{\citenamefont {Prosen}(2012)}]{Prosen2012}%
  \BibitemOpen
  \bibfield  {author} {\bibinfo {author} {\bibfnamefont {T.}~\bibnamefont
  {Prosen}},\ }\href {https://doi.org/10.1103/PhysRevLett.109.090404}
  {\bibfield  {journal} {\bibinfo  {journal} {Phys. Rev. Lett.}\ }\textbf
  {\bibinfo {volume} {109}},\ \bibinfo {pages} {090404} (\bibinfo {year}
  {2012})}\BibitemShut {NoStop}%
\bibitem [{\citenamefont {Huber}\ \emph {et~al.}(2020)\citenamefont {Huber},
  \citenamefont {Kirton}, \citenamefont {Rotter},\ and\ \citenamefont
  {Rabl}}]{Huber2020}%
  \BibitemOpen
  \bibfield  {author} {\bibinfo {author} {\bibfnamefont {J.}~\bibnamefont
  {Huber}}, \bibinfo {author} {\bibfnamefont {P.}~\bibnamefont {Kirton}},
  \bibinfo {author} {\bibfnamefont {S.}~\bibnamefont {Rotter}},\ and\ \bibinfo
  {author} {\bibfnamefont {P.}~\bibnamefont {Rabl}},\ }\href
  {https://doi.org/10.21468/SciPostPhys.9.4.052} {\bibfield  {journal}
  {\bibinfo  {journal} {SciPost Phys.}\ }\textbf {\bibinfo {volume} {9}},\
  \bibinfo {pages} {052} (\bibinfo {year} {2020})}\BibitemShut {NoStop}%
\bibitem [{\citenamefont {Nakanishi}\ and\ \citenamefont
  {Sasamoto}(2022)}]{Nakanishi2022}%
  \BibitemOpen
  \bibfield  {author} {\bibinfo {author} {\bibfnamefont {Y.}~\bibnamefont
  {Nakanishi}}\ and\ \bibinfo {author} {\bibfnamefont {T.}~\bibnamefont
  {Sasamoto}},\ }\href {https://doi.org/10.1103/PhysRevA.105.022219} {\bibfield
   {journal} {\bibinfo  {journal} {Phys. Rev. A}\ }\textbf {\bibinfo {volume}
  {105}},\ \bibinfo {pages} {022219} (\bibinfo {year} {2022})}\BibitemShut
  {NoStop}%
\bibitem [{\citenamefont {Nakanishi}\ and\ \citenamefont
  {Sasamoto}(2023)}]{Nakanishi2023}%
  \BibitemOpen
  \bibfield  {author} {\bibinfo {author} {\bibfnamefont {Y.}~\bibnamefont
  {Nakanishi}}\ and\ \bibinfo {author} {\bibfnamefont {T.}~\bibnamefont
  {Sasamoto}},\ }\href {https://doi.org/10.1103/PhysRevA.107.L010201}
  {\bibfield  {journal} {\bibinfo  {journal} {Phys. Rev. A}\ }\textbf {\bibinfo
  {volume} {107}},\ \bibinfo {pages} {L010201} (\bibinfo {year}
  {2023})}\BibitemShut {NoStop}%
\bibitem [{\citenamefont {Landig}\ \emph {et~al.}(2016)\citenamefont {Landig},
  \citenamefont {Hruby}, \citenamefont {Dogra}, \citenamefont {Landini},
  \citenamefont {Mottl}, \citenamefont {Donner},\ and\ \citenamefont
  {Esslinger}}]{Esslinger2}%
  \BibitemOpen
  \bibfield  {author} {\bibinfo {author} {\bibfnamefont {R.}~\bibnamefont
  {Landig}}, \bibinfo {author} {\bibfnamefont {L.}~\bibnamefont {Hruby}},
  \bibinfo {author} {\bibfnamefont {N.}~\bibnamefont {Dogra}}, \bibinfo
  {author} {\bibfnamefont {M.}~\bibnamefont {Landini}}, \bibinfo {author}
  {\bibfnamefont {R.}~\bibnamefont {Mottl}}, \bibinfo {author} {\bibfnamefont
  {T.}~\bibnamefont {Donner}},\ and\ \bibinfo {author} {\bibfnamefont
  {T.}~\bibnamefont {Esslinger}},\ }\href {https://doi.org/10.1038/nature17409}
  {\bibfield  {journal} {\bibinfo  {journal} {Nature}\ }\textbf {\bibinfo
  {volume} {532}},\ \bibinfo {pages} {476} (\bibinfo {year}
  {2016})}\BibitemShut {NoStop}%
\bibitem [{\citenamefont {Chitra}\ and\ \citenamefont
  {Zilberberg}(2015)}]{Chitra2015}%
  \BibitemOpen
  \bibfield  {author} {\bibinfo {author} {\bibfnamefont {R.}~\bibnamefont
  {Chitra}}\ and\ \bibinfo {author} {\bibfnamefont {O.}~\bibnamefont
  {Zilberberg}},\ }\href {https://doi.org/10.1103/PhysRevA.92.023815}
  {\bibfield  {journal} {\bibinfo  {journal} {Physical Review A}\ }\textbf
  {\bibinfo {volume} {92}},\ \bibinfo {pages} {023815} (\bibinfo {year}
  {2015})}\BibitemShut {NoStop}%
\bibitem [{\citenamefont {Zhu}\ \emph {et~al.}(2019)\citenamefont {Zhu},
  \citenamefont {Marino}, \citenamefont {Yao}, \citenamefont {Lukin},\ and\
  \citenamefont {Demler}}]{Zhu2019}%
  \BibitemOpen
  \bibfield  {author} {\bibinfo {author} {\bibfnamefont {B.}~\bibnamefont
  {Zhu}}, \bibinfo {author} {\bibfnamefont {J.}~\bibnamefont {Marino}},
  \bibinfo {author} {\bibfnamefont {N.~Y.}\ \bibnamefont {Yao}}, \bibinfo
  {author} {\bibfnamefont {M.~D.}\ \bibnamefont {Lukin}},\ and\ \bibinfo
  {author} {\bibfnamefont {E.~A.}\ \bibnamefont {Demler}},\ }\href
  {https://doi.org/10.1088/1367-2630/ab2afe} {\bibfield  {journal} {\bibinfo
  {journal} {New Journal of Physics}\ }\textbf {\bibinfo {volume} {21}},\
  \bibinfo {pages} {073028} (\bibinfo {year} {2019})}\BibitemShut {NoStop}%
\bibitem [{\citenamefont {Chelpanova}\ \emph {et~al.}(2021)\citenamefont
  {Chelpanova}, \citenamefont {Lerose}, \citenamefont {Zhang}, \citenamefont
  {Carusotto}, \citenamefont {Tserkovnyak},\ and\ \citenamefont
  {Marino}}]{Chelpanova2021}%
  \BibitemOpen
  \bibfield  {author} {\bibinfo {author} {\bibfnamefont {O.}~\bibnamefont
  {Chelpanova}}, \bibinfo {author} {\bibfnamefont {A.}~\bibnamefont {Lerose}},
  \bibinfo {author} {\bibfnamefont {S.}~\bibnamefont {Zhang}}, \bibinfo
  {author} {\bibfnamefont {I.}~\bibnamefont {Carusotto}}, \bibinfo {author}
  {\bibfnamefont {Y.}~\bibnamefont {Tserkovnyak}},\ and\ \bibinfo {author}
  {\bibfnamefont {J.}~\bibnamefont {Marino}},\ }\href
  {https://doi.org/10.48550/arXiv.2112.04509} {\bibfield  {journal} {\bibinfo
  {journal} {arXiv.2112.04509}\ } (\bibinfo {year} {2021})}\BibitemShut
  {NoStop}%
\end{thebibliography}%

%\end{document}

\newpage
\clearpage
\appendix
	
\setcounter{figure}{0}
\makeatletter 
\renewcommand{\thefigure}{S\arabic{figure}}
	
\newcounter{defcounter}
\setcounter{defcounter}{0}
	
\newenvironment{myequation}
{
\addtocounter{equation}{-1}
\refstepcounter{defcounter}
\renewcommand\theequation{S\thedefcounter}
\align
}
{
\endalign
}
\setcounter{page}{1}
	
\begin{widetext}
\begin{center}
{\fontsize{12}{12}\selectfont
\textbf{Supplemental Material for ``\papertitle''\\[5mm]}}
%{\normalsize \authornames\\[1mm]}
%{\fontsize{9}{9}\selectfont  
%\textit{\tcm}}
\end{center}
\normalsize
\end{widetext}

This Supplemental Material consists of two sections. In the first section we provide further details on the origin of photon-mediated non-reciprocal interaction and report the explicit expressions of the effective non-reciprocal dynamics obtained from Eq.~\eqref{eq:er2}. In the second section we provide additional information on the non-stationary dynamics described in the main text, with reference to Fig.~\ref{Time_Dynamics}.

\section{Origin of non-reciprocity}
In this section we discuss the emergence of non-reciprocity in the NRDM.
\begin{figure}[t]
\centering
\includegraphics[width=\columnwidth]{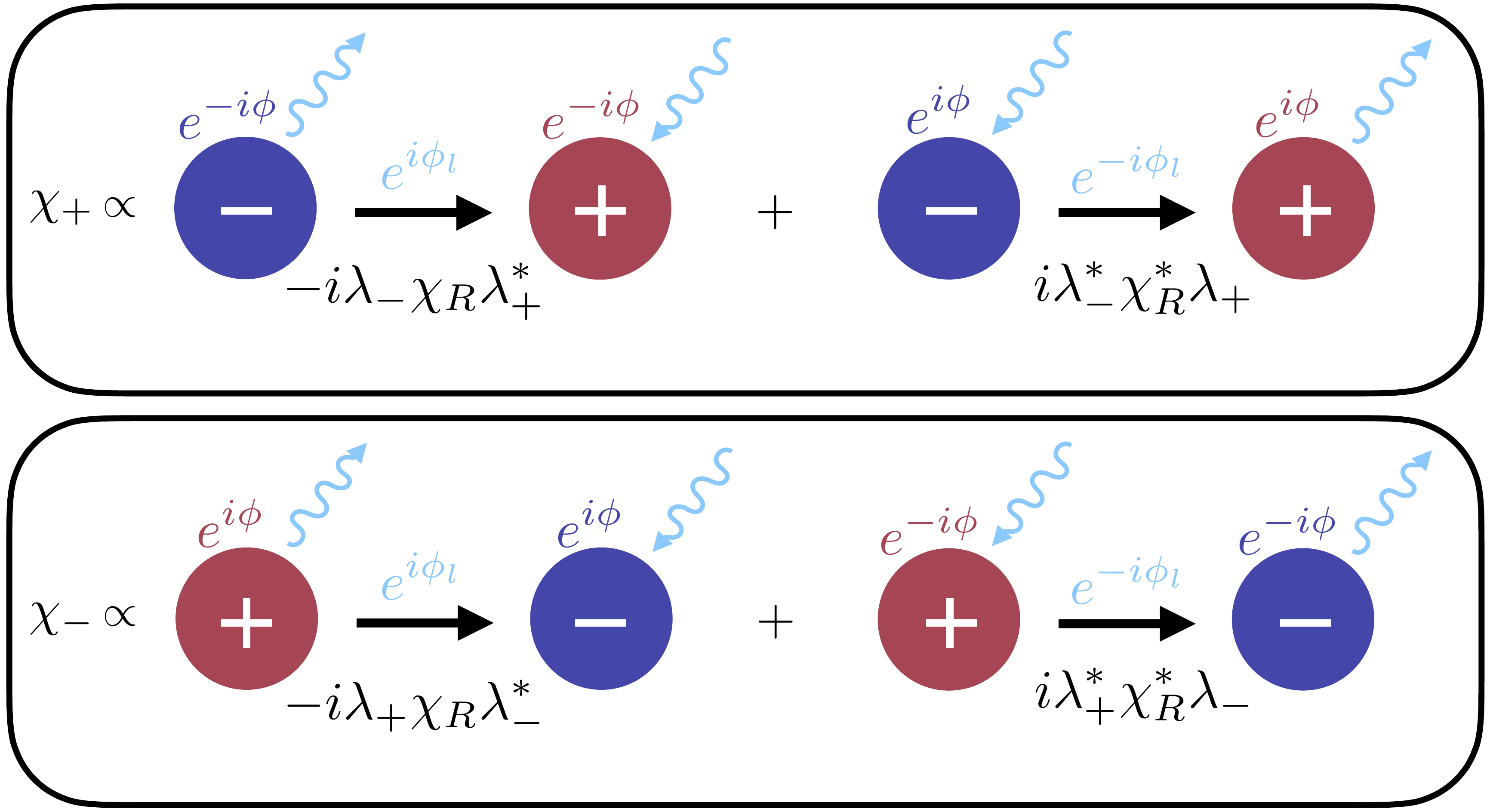}
\caption{\label{Non-rec-scheme}\textbf{Non-reciprocal interactions mediated by photon scattering}. The spin species are denoted by red and blue circles. Light blue arrows denote interactions with an incoming or outgoing photon.  Non-reciprocal inter-species couplings $\chi_{\pm}$ result from the interference of two different scattering processes. The phase factors during the spin-photon interactions follow from the light-matter couplings $\lambda_{\pm} = \lambda e^{\pm i\phi}$, and the photon phase shift $\phi_{l} = \arctan{(2\omega_{l}/\kappa)}$ from the light field response $\chi_{R} = |\chi_{R}| e^{i\phi_{l}}$.}
\end{figure}
In its simplest form, the origin of non-reciprocal interactions can be unveiled by considering the limit in which the light field can be adiabatically eliminated from the dynamics $(\omega_{l},\kappa) \gg (\omega_{0},\delta,\Gamma_{\downarrow},\lambda)$. Setting the light field to its steady-state value $\alpha_{SS}$, we obtain
\begin{equation}
    \label{eq:erA001}
    \alpha_{SS} = \frac{i\sqrt{N}}{2} \chi_{R} \left(\lambda_{+} s_{x,+} + \lambda_{-} s_{x,-} \right),
\end{equation}
where we introduced the notation $\lambda_{\pm} = \lambda e^{\pm i\phi}$ for the coupling between the light field and the species '$\pm$', and $\chi_{R} = (i\omega_{l} - \kappa/2)^{-1}$ the light field response function. Upon inserting this expression in the equations of motion of the spin species we are left with
\begin{align}
\label{eq:erA002}
&\dot{s}_{x,\pm} = -( \omega_0 \pm \delta) s_{y,\pm} - \frac{\Gamma_{\downarrow}}{2} s_{x,\pm} \nonumber \\
&\dot{s}_{y,\pm} = ( \omega_0 \pm \delta) s_{x,\pm} - \frac{\Gamma_{\downarrow}}{2} s_{y,\pm} +\xi s_{x,\pm}s_{z,\pm} - \chi_{\pm}s_{x,\mp}s_{z,\pm} \nonumber \\
&\dot{s}_{z,\pm} = -\Gamma_{\downarrow} (s_{z,\pm} + 1) + \xi s_{x,\pm}s_{y,\pm} - \chi_{\pm}s_{x,\mp}s_{y,\pm}.
\end{align}
From these equations, we observe how the light field mediates intra-species interactions, with strength $\xi = \lambda^{2} \textrm{Im}[\chi_{R}]$, and inter-species interactions with coupling strength
\begin{equation}
\label{eq:erA003}
\chi_{\pm} = -\frac{i}{2} \left( \lambda_{\pm}^{*} \chi_{R} \lambda_{\mp} - \lambda_{\pm} \chi_{R} \lambda^{*}_{\mp} \right).
\end{equation}
For $\phi \neq 0$ and $\kappa \neq 0$, the interactions between the spin species are non-reciprocal $\chi_{+} \neq \chi_{-}$. This becomes evident when recasting the couplings as $\chi_{\pm} = \lambda^{2} |\chi_{R}| \sin{(\phi_{l} \mp 2 \phi)}$, where the phase shift induced by the light field is $\phi_{l} = \arctan{(2\omega_{l}/\kappa)}$. For $\phi = 0$, the couplings become trivially symmetric, as expected for the regular open Dicke model. Analogously, in the absence of dissipation $(\kappa = 0)$, $\phi_{l} \rightarrow \frac{\pi}{2}$ and $\sin{(\frac{\pi}{2} \mp 2\phi)} = \cos{(2\phi)}$, thus also retrieving symmetrical couplings.

Physically, non-reciprocity can be understood as an asymmetrical interference of photon-mediated scattering processes between the two spin species.
In this picture, the couplings $\chi_{\pm}$ correspond to the total scattering amplitude, with each term in \eqref{eq:erA003} being associated with a photon scattering process from one species to the other. This is illustrated in Fig.~\ref{Non-rec-scheme}.

%{\color{blue}
\section{Stability analysis}

In this section, we outline the main steps behind the linear stability analysis used to construct the phase diagrams in Fig.~\ref{PD-Spec} from the main text.

First, the steady-state value of the system variables are obtained by setting the equations of motion (\ref{eq:er2}) to zero, yielding a system of non-linear algebraic equations which in general needs to be solved numerically.
Next, we analyze the behavior of linear fluctuations around the steady-states, which determine the stability of the solutions and in turn the stability of a specific phase. 
This is done through the matrix of coefficients that dictates the evolution of the fluctuations, commonly known as the \textit{dynamical matrix}. 
Specifically, we study the eigenvalues of the dynamical matrix, which we refer to as \textit{dynamical spectrum}. The imaginary part of these eigenvalues describes the oscillatory dynamics of the fluctuations, while the real part reveals whether the solution is stable or not. 
If the real part is negative, fluctuations will decay and vanish in the long-time limit. 
On the contrary, if these are positive, fluctuations will grow exponentially and the solution will be rendered unstable.

Below, we present the two different analyses corresponding to the limit in which the light field is adiabatically eliminated and when the full system is considered. While the same algebraic equations need to be solved in both cases, the main difference between them is the dynamical matrix that is used to investigate the stability of the solutions, which will either include or not the fluctuations of the light field. For simplicity, we focus on the stability of the normal phase, where either analytical expressions can be derived or the dynamical matrix acquires a simpler form. Outside the normal phase, the analysis can be easily carried out numerically.

\subsection{Adiabatic elimination regime}

Upon elimination of the photonic degrees of freedom, the dynamics of the system are given by Eqs.~\eqref{eq:erA002}. Performing a linear expansion of the spin degrees of freedom around their steady state value $s_{\alpha,\pm}(t) \simeq s_{\alpha,\pm}^{SS} + \delta s_{\alpha,\pm}(t)$, which in the normal phase read $(s_{x,\pm}^{SS},s_{y,\pm}^{SS},s_{z,\pm}^{SS}) = (0,0,-1)$, we obtain the linearized equations of motion
\begin{widetext}
\begin{equation}
    \begin{pmatrix}
    \delta \dot{s}_{x,+} \\
    \delta \dot{s}_{y,+} \\
    \delta \dot{s}_{z,+} \\
    \delta \dot{s}_{x,-} \\
    \delta \dot{s}_{y,-} \\
    \delta \dot{s}_{z,-} 
    \end{pmatrix}
    =
    \begin{pmatrix}
    -\frac{\Gamma_{\downarrow}}{2} & -(\omega_{0} + \delta) & 0 & 0 & 0 & 0 \\
    \omega_{0} + \delta + \xi & -\frac{\Gamma_{\downarrow}}{2} & 0 & -\chi_{+} & 0 & 0 \\
    0 & 0 & -\Gamma_{\downarrow} & 0 & 0 & 0\\
    0 & 0 & 0 & -\frac{\Gamma_{\downarrow}}{2} & -(\omega_{0} - \delta) & 0  \\
    -\chi_{-} & 0 & 0 & \omega_{0} - \delta + \xi & -\frac{\Gamma_{\downarrow}}{2} & 0 \\
    0 & 0 & 0 & 0 & 0 & -\Gamma_{\downarrow} 
    \end{pmatrix}
    \begin{pmatrix}
    \delta s_{x,+} \\
    \delta s_{y,+} \\
    \delta s_{z,+} \\
    \delta s_{x,-} \\
    \delta s_{y,-} \\
    \delta s_{z,-}.
    \end{pmatrix},
\end{equation}
\end{widetext}
where the coefficients $\xi$ and $\chi_{\pm}$ are the same as in the previous section.
This dynamical matrix can be diagonalized analytically, yielding the dynamical spectrum
\begin{widetext}
\begin{equation}
\label{eq:erB001}
    \eta_{\pm,\pm} = -\frac{\Gamma_{\downarrow}}{2} \pm \sqrt{-\omega_{0}(\omega_{0} + \xi) - \delta^{2} \pm \sqrt{\delta^2(2\omega_{0}+\xi)^{2} + (\omega_{0}^{2}-\delta^{2})\chi_{+}\chi_{-}}},
\end{equation}
\end{widetext}
together with an extra pair of degenerate eigenvalues $\eta_{1,2} = -\Gamma_{\downarrow}$ resulting from fluctuations along the $z$ components.
In Fig.~\ref{Dyn-Spec}, we show in blue the real part of the dynamical spectrum, resulting from Eq.~\eqref{eq:erB001}, when crossing the NP-DP boundary for (a) $\delta = \Gamma_{\downarrow} = 0$, (b) $\delta = 0$ and $\Gamma_{\downarrow} \neq 0$, and (c) $\delta \neq 0$ and $\Gamma_{\downarrow} = 0$.
Importantly, for all cases we observe the presence of exceptional points, characteristic of the effective non-reciprocal interactions following the adiabatic elimination of the light field.
Note that (a) is equivalent to Fig.~\ref{PD-Spec}(b) in the main text.
% In this case, the NP-DP transition occurs when $\chi_{+}\chi_{-} = 0$, even for infinitesimal values of $\lambda$, yielding the lines of exceptional points present in the phase diagram shown in Fig.~\ref{PD-Spec}(a).
% For (b) and (c), the presence spin decay and frequency imbalance add robustness to the NP. 
In (b), spin decay shifts the spectrum, resulting in the transition taking place beyond the exceptional points, and only when $\lambda$ is large enough to overcome the $\Gamma_{\downarrow}/2$ gap.
%, thus adding robustness to the NP. 
For (c), despite the spectrum having the same structure as (a), the condition $\chi_{+}\chi_{-} < 0$ is in general not enough for the transition to occur, meaning that the DP will no longer emerge at arbitrarily small $\lambda$.
%, leading to frequency imbalance stabilizing the NP in a larger region.
As a result, the presence of either spin decay or frequency imbalance will stabilize the NP. 

\subsection{Full system}

\begin{figure}[t]
\centering
\includegraphics[width=\columnwidth]{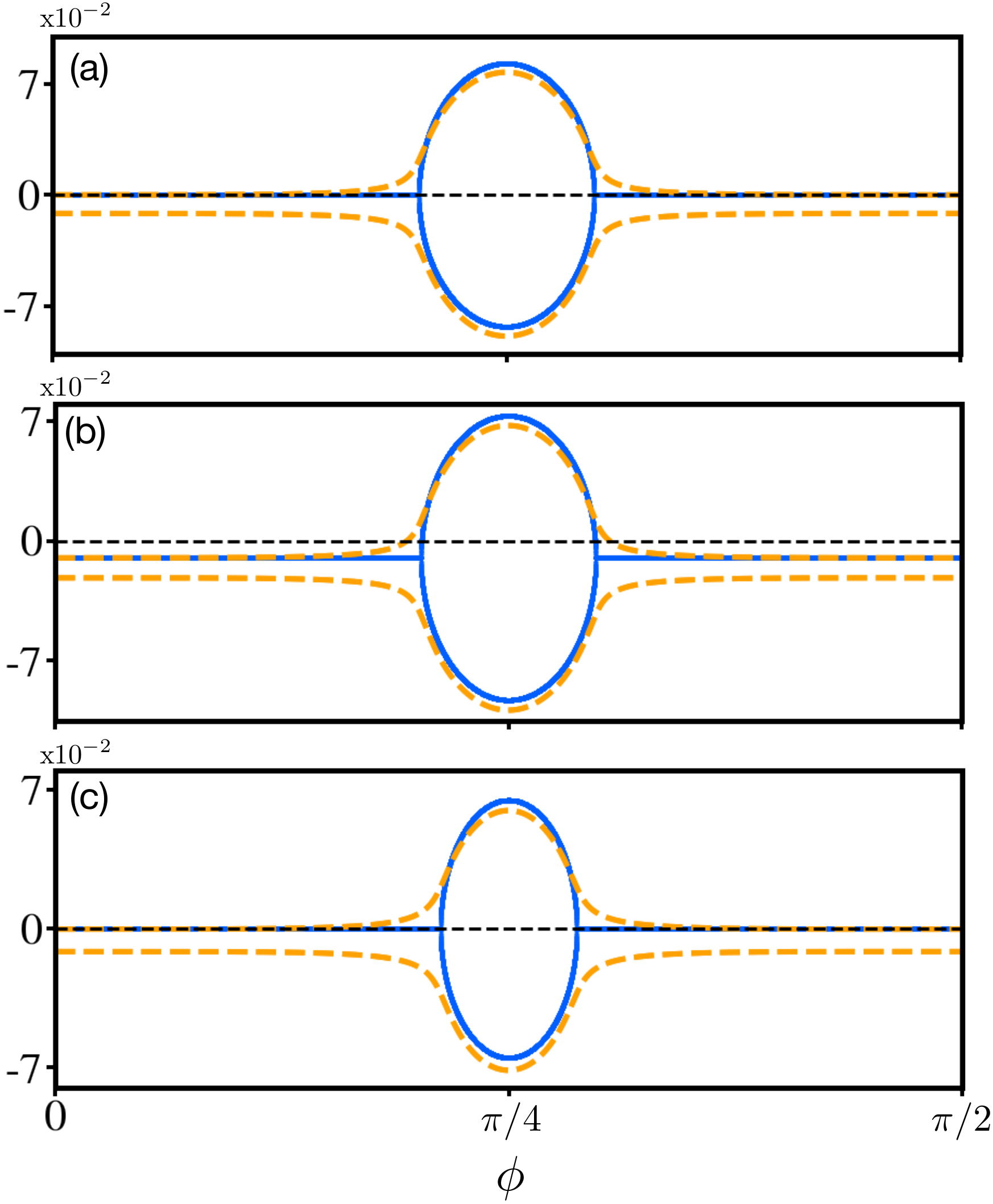}
\caption{\label{Dyn-Spec}\textbf{Real part of the dynamical spectrum} for (a) $\delta = \Gamma_{\downarrow}=0$, (b) $\delta = 0$, $\Gamma_{\downarrow} = 0.02\omega_{0}$ and (c) $\delta = 0.05\omega_{0}$, $\Gamma_{\downarrow} = 0$. Blue lines correspond to the spectrum in the adiabatically eliminated case, see Eq.~\eqref{eq:erB001}, and gold dashed lines to the spectrum of the full system, obtained from diagonalizing numerically the dynamical matrix in \eqref{eq:erB002}. Parameters: $\omega_{l}  = 20\omega_{0}$, $\kappa = 12.5\omega_{0}$ and $\lambda = 3 \omega_{0}$.}
\end{figure}

We now consider the fluctuation dynamics of the full system, which is obtained by also considering the linear fluctuations of the light field $\alpha(t) = \alpha^{SS} + \delta \alpha(t)$, where $\alpha^{SS} = 0$ in the normal phase. The dynamics is given by
\begin{widetext}
\begin{equation}
\label{eq:erB002}    
    \begin{pmatrix}
    \delta \dot{\alpha} \\
    \delta \dot{\alpha}^{*} \\
    \delta \dot{s}_{x,+} \\
    \delta \dot{s}_{y,+} \\
    \delta \dot{s}_{z,+} \\
    \delta \dot{s}_{x,-} \\
    \delta \dot{s}_{y,-} \\
    \delta \dot{s}_{z,-} 
    \end{pmatrix}
    =
    \begin{pmatrix}
    -i\omega_{l} - \frac{\kappa}{2} & 0 & -i\frac{\lambda\sqrt{N}}{2}e^{i\phi} & 0 & 0 & -i\frac{\lambda\sqrt{N}}{2}e^{-i\phi} & 0 & 0 \\
    0 & i\omega_{l} - \frac{\kappa}{2}  & i\frac{\lambda\sqrt{N}}{2}e^{-i\phi} & 0 & 0 & i\frac{\lambda\sqrt{N}}{2}e^{i\phi} & 0 & 0 \\
    0 & 0 & -\frac{\Gamma_{\downarrow}}{2} & -(\omega_{0} + \delta) & 0 & 0 & 0 & 0 \\
    \frac{\lambda}{\sqrt{N}}e^{-i\phi} & \frac{\lambda}{\sqrt{N}}e^{i\phi}  & \omega_{0} + \delta & -\frac{\Gamma_{\downarrow}}{2} & 0 & 0 & 0 & 0 \\
    0 & 0 & 0 & 0 & -\Gamma_{\downarrow} & 0 & 0 & 0\\
    0 & 0 & 0 & 0 & 0 & -\frac{\Gamma_{\downarrow}}{2} & -(\omega_{0} - \delta) & 0  \\
    \frac{\lambda}{\sqrt{N}}e^{i\phi} & \frac{\lambda}{\sqrt{N}}e^{-i\phi}  & 0 & 0 & 0 & \omega_{0} - \delta & -\frac{\Gamma_{\downarrow}}{2} & 0 \\
    0 & 0 & 0 & 0 & 0 & 0 & 0 & -\Gamma_{\downarrow} 
    \end{pmatrix}
    \begin{pmatrix}
    \delta \alpha \\
    \delta \alpha^{*} \\
    \delta s_{x,+} \\
    \delta s_{y,+} \\
    \delta s_{z,+} \\
    \delta s_{x,-} \\
    \delta s_{y,-} \\
    \delta s_{z,-}.
    \end{pmatrix}.
\end{equation}
\end{widetext}
In this case the spectrum cannot be computed analytically, but it can easily be obtained numerically.
The real part of the spectrum for the full dynamics is shown in gold in Fig.~\ref{Dyn-Spec}.
%We observe how all the exceptional points disappear as a result of the light field fluctuation dynamics.
For (a), this result corresponds to the gold dashed line shown in Fig.~\ref{PD-Spec}(b), where we observe how presence the light field fluctuations remove the exceptional points from the spectrum and renders the normal phase unstable for all $\phi$, except $\phi = 0, \frac{\pi}{2}$, where interactions are reciprocal.
As expected, for (b) and (c), the light field dynamics softens the spectrum and lifts the exceptional points.
Note that these determine the shape of the phase diagrams shown in Figs.~\ref{PD-Spec}(c) and (d) for $\lambda = 3\omega_{0}$.

\section{Details regarding steady-state dynamics in the DP}

In this section, we provide more details regarding the analysis of the system in the non-stationary steady-state regime inside the DP. For concreteness, we focus on the case $\nobreak{\delta = \Gamma_{\downarrow}=0}$.
Our starting point is the numerical evidence that the $z$ component of one of the two species approaches a stationary value in the long time limit, as shown in Fig.~\ref{Time_Dynamics}(a), (b) of the main text. By importing this knowledge in Eqs.~\eqref{eq:er2} and setting $\dot{s}_{z,-} = 0$, we obtain $\lambda \left( \alpha e^{ i\phi} + \alpha^{*} e^{- i\phi}  \right)s_{y,-} = 0$. From numerical observations we further know that both $\alpha$ [cf.~Fig.~\ref{Spectrum}(a)] and $s_{y,-}$ (not shown) remain oscillatory in the steady-state regime. Therefore, by using the decomposition $\alpha (t) = A(t)e^{i\varphi(t)}$, we are left with the condition $\cos{(\varphi(t) + \phi)} = 0$, yielding $\varphi = \frac{\pi}{2} - \phi$. Note that setting $\dot{s}_{z,+} = 0$ instead, would lead to $\varphi = \frac{\pi}{2} + \phi$. The phase locking results in the light field decoupling from the $``-"$ spin species, leading to free evolution of $s_{(x,y),-}$.

From the stationarity of the phase, we can obtain an analytical expression for the amplitude using the equation of motion of  $\varphi$, which reads
\begin{equation}
\label{eq:erSM2001}
A \dot{\varphi}  = -\omega_{l} A - \frac{\lambda \sqrt{N}}{2} (s_{x,+} \cos{(\phi - \varphi)} + s_{x,-} \cos{(\phi + \varphi)}) .
\end{equation}
Setting $\dot{\varphi} =  0$ and using the phase locking condition, we arrive at $A(t) = -\frac{\lambda\sqrt{N} \sin{(2\phi)}}{2\omega_{l}} s_{x,+}(t)$, showing how phase locking leads to the enslaving of the photons to the spins. Substituting this into the equations of motion of the $``+"$ spins, their dynamics is given by the system of equations
\begin{align}
\label{eq:erSM2002}
&\dot{s}_{x,+} = -\omega_{0} s_{y,+} \nonumber \\
&\dot{s}_{y,+} = \omega_{0} s_{x,+} + \frac{\lambda^{2}}{2\omega_{l}} \sin^{2}{(2\phi)} s_{x,+} s_{z,+} \nonumber \\
&\dot{s}_{z,+} = -\frac{\lambda^{2}}{2\omega_{l}} \sin^{2}{(2\phi)} s_{x,+} s_{y,+}.
\end{align}
The interactions with the phase-locked light field then result in more complex dynamics for the $``+"$ spins, leading to the frequency doubling of $s_{z,+}$.

\end{document}